\documentclass[twocolumn,usenatbib,useAMS,fleqn,usenames,dvipsnames]{aastex63}
\usepackage{amsmath}
\hypersetup{linkcolor=Aquamarine,citecolor=MidnightBlue,filecolor=cyan,urlcolor=MidnightBlue}
\graphicspath{{./}{figures_color_coded/}{figures_lines/}}
\received{}
\revised{}
\accepted{}
\submitjournal{ApJ}
\shorttitle{The simulated CGM}
\shortauthors{Fielding \& the SMAUG team}

\newcommand{\Msun}{\rm{M_{\odot}}}
\newcommand{\Mhalo}{\rm{M_{200c}}}
\newcommand{\rvir}{r_{\rm 200c}}
\newcommand{\Tvir}{T_{\rm 200c}}
\newcommand{\nvir}{n_{\rm 200c}}
\newcommand{\Pvir}{P_{\rm 200c}}
\newcommand{\Kvir}{K_{\rm 200c}}
\newcommand{\vvir}{v_{\rm 200c}}
\newcommand{\mump}{\mu m_{\rm p}}
\newcommand{\kb}{k_{\rm B}}
\newcommand{\fb}{f_{\rm b}}

\defcitealias{Fielding+17}{F17}
\defcitealias{Su+20}{S20}
\defcitealias{Li+20a}{L20a}
\defcitealias{Li+20b}{L20b}
\defcitealias{Joung+12}{J12}

\begin{document}

\title{First results from SMAUG: Uncovering the Origin of the Multiphase Circumgalactic Medium with a Comparative Analysis of Idealized and Cosmological Simulations}

\correspondingauthor{Drummond B. Fielding}
\email{drummondfielding@gmail.com}

\author[0000-0003-3806-8548]{Drummond B. Fielding}
\affiliation{Center for Computational Astrophysics, Flatiron Institute, 162 5th Ave, New York, NY 10010, USA}

\author[0000-0002-8710-9206]{Stephanie Tonnesen}
\affiliation{Center for Computational Astrophysics, Flatiron Institute, 162 5th Ave, New York, NY 10010, USA}

\author[0000-0002-0112-7690]{Daniel DeFelippis} 
\affiliation{Department of Astronomy, Columbia University, 550 W 120th Street, New York, NY 10027, USA}

\author[0000-0003-0773-582X]{Miao Li}
\affiliation{Center for Computational Astrophysics, Flatiron Institute, 162 5th Ave, New York, NY 10010, USA}

\author[0000-0003-1598-0083]{Kung-Yi Su}
\affiliation{Center for Computational Astrophysics, Flatiron Institute, 162 5th Ave, New York, NY 10010, USA}

\author[0000-0003-2630-9228]{Greg L. Bryan}
\affiliation{Department of Astronomy, Columbia University, 550 W 120th Street, New York, NY 10027, USA}
\affiliation{Center for Computational Astrophysics, Flatiron Institute, 162 5th Ave, New York, NY 10010, USA}

\author[0000-0003-2896-3725]{Chang-Goo Kim}
\affiliation{Department of Astrophysical Sciences, Princeton University, 4 Ivy Lane, Princeton, NJ 08544, USA}
\affiliation{Center for Computational Astrophysics, Flatiron Institute, 162 5th Ave, New York, NY 10010, USA}

\author[0000-0002-1975-4449]{John C. Forbes}
\affiliation{Center for Computational Astrophysics, Flatiron Institute, 162 5th Ave, New York, NY 10010, USA}

\author[0000-0003-2835-8533]{Rachel S. Somerville}
\affiliation{Center for Computational Astrophysics, Flatiron Institute, 162 5th Ave, New York, NY 10010, USA}
\affiliation{Department of Physics and Astronomy, Rutgers University, 136 Frelinghuysen Rd, Piscataway, NJ 08854, USA}

\author[0000-0001-5846-0411]{Nicholas Battaglia}
\affiliation{Department of Astronomy, Cornell University, Ithaca, NY 14853 USA}

\author[0000-0001-9735-7484]{Evan E. Schneider}
\affiliation{Department of Astrophysical Sciences, Princeton University, 4 Ivy Lane, Princeton, NJ 08544, USA}

\author[0000-0001-5262-6150]{Yuan Li}
\affiliation{Department of Astronomy, University of California, Berkeley, CA 94720, USA}

\author[0000-0002-8131-6378]{Ena Choi}
\affil{Quantum Universe Center, Korea Institute for Advanced Study, Hoegiro 85, Seoul 02455, Korea}

\author[0000-0003-4073-3236]{Christopher C. Hayward}
\affiliation{Center for Computational Astrophysics, Flatiron Institute, 162 5th Ave, New York, NY 10010, USA}

\author[0000-0001-6950-1629]{Lars Hernquist}
\affiliation{Harvard-Smithsonian Center for Astrophysics, 60 Garden Street, Cambridge, MA 02138, USA}

\begin{abstract}

We examine the properties of the circumgalactic medium (CGM) at low redshift in a range of simulated Milky Way mass halos. The sample is comprised of seven idealized simulations, an adaptive mesh refinement cosmological zoom-in simulation, and two groups of 50 halos with star forming or quiescent galaxies taken from the IllustrisTNG100 simulation. The simulations have very different setups, resolution, and feedback models, but are analyzed in a uniform manner. By comparing median radial profiles and mass distributions of CGM properties, we isolate key similarities and differences. In doing so, we advance the efforts of the SMAUG (Simulating Multiscale Astrophysics to Understand Galaxies) project that aims to understand the inherently multiscale galaxy formation process. In the cosmological simulations, the CGM exhibits nearly flat temperature distributions, and broad pressure and radial velocity distributions. In the idealized simulations, similar distributions are found in the inner CGM ($\lesssim 0.5 \, \rvir$) when strong galactic feedback models are employed, but the outer CGM ($\gtrsim 0.5 \, \rvir$) has a much less prominent cold phase, and narrower pressure and velocity distributions even in models with strong feedback. 
This comparative analysis demonstrates the dominant role feedback plays in shaping the inner CGM and the increased importance of cosmological effects, such as nonspherical accretion and satellite galaxies, in the outer CGM. Furthermore, our findings highlight that while cosmological simulations are required to capture the multiphase structure of the CGM at large radii, idealized simulations provide a robust framework to study how galactic feedback interacts with the inner CGM and thereby provide a reliable avenue to constrain feedback prescriptions.  
\end{abstract}


\section{Introduction} \label{sec:intro}

The flow of gas into and out of the CGM regulates galaxy growth over cosmic time. Recent observations have painted a tantalizing picture of these important flows and the relationship between CGM and galaxy properties \citep[see][for a recent review]{Tumlinson+17}. 
Observations across a range of wavelengths including X-rays, UV, and optical have demonstrated that there exists copious, highly enriched \citep{Werk+14, Prochaska+17, Lehner+19} gas at a broad range of ionization states (and, therefore, a broad range of temperatures and densities)\citep{Chen+10, Prochaska+11, Burchett+19} with diverse kinematic properties \citep{Werk+16, Nielsen+17, Rudie+19} in the CGM around both star forming and quiescent galaxies at low and high redshifts \citep{Steidel+10, Tumlinson+11, Thom+12, Bordoloi+14, Borthakur+15, Burchett+16, Johnson+17, Zahedy+19}.
To date, however, most CGM observations are from absorption line studies that usually provide only a single sight line per galaxy and require modeling that is fraught with many degeneracies to extract physical properties. 

Given the challenges in observing the CGM, numerical simulations have played an outsized role in filling in the gaps of our knowledge and in shaping our view of the physical processes at play. There are two main approaches to simulating the CGM that can be broadly categorized as cosmological and idealized. 

Cosmological simulations entail simulating portions of the universe starting from high redshift with cosmological initial conditions down to present times. These simulations often include many physical processes related to galaxy formation in the form of sub-grid models. This approach of including as many processes as possible has the benefit of a high degree of physical realism, but comes at the cost of high complexity and difficulty in isolating the dominant physical process behind specific properties. Moreover, in many cases the sub-grid models are not physically motivated and are instead tuned to match specific observed properties of galaxies, which limits the predictive ability of these types of simulations. Large volume cosmological simulations contain numerous galaxies which allows for statistical studies of different populations. This, however, often comes at the cost of coarser spatial resolution, so it is not possible to accurately resolve the dynamics and evolution of small cold clouds in the CGM. 
Cosmological simulations have been used to study CGM observational characteristics \citep[e.g.,][]{Hummels+13,CAFG+15,Liang+16,Oppenheimer+16,Hafen+17,Gutcke+17,Nelson+18} as well as the physical nature of CGM gas \citep[e.g.,][]{Oppenheimer18, Hafen+19a, Hafen+19b, Ji+20,DeFelippis+20}. 

In addition, different cosmological simulations have vastly different CGM properties \citep[e.g.,][]{Davies+19}. The CGM is an important point of comparison, because while many simulations are tuned to agree with global galaxy properties, the CGM is not directly modeled. Indeed, V. Pandya et al. (2020 in preparation) compare the FIRE simulations to SAMs with similar galaxy properties and find that they produce dramatically different CGMs.  Until the impact of different physical processes and sub-grid models on the CGM is well-understood we cannot leverage the constraining power of halo gas on cosmological simulations.

Idealized simulations, on the other hand, simulate individual galaxy halos removed from the cosmological context. These simulations include a selection of hand-picked ingredients. As a result the simulations are easier to interpret, but lack physical realism. They are generally less computationally expensive than cosmological simulations and can be run with higher resolution. Ingredients can be added incrementally to isolate their impact on the CGM structure and evolution. Global idealized CGM simulations have been used to study the bulk properties and phase structure of the CGM and how they depend on properties of the galaxy and halo \citep[e.g.,][]{Fielding+17, Su+20, Li+20a, Stern+19, Stern+20, Lochhaas+20}.

Even when simulations nominally include the same physical processes, their implementation within a sub-grid model can vary dramatically.  For example, star formation feedback may increase the thermal and kinetic energy of the surrounding gas in varying ratios \citep[e.g.,][]{Hopkins+18}.  Supernova feedback may be implemented instantaneously or have a lag-time of tens of Myr \citep[e.g.,][]{Stinson+06,EAGLE}.  In some simulations supernovae may produce ``wind particles" that are decoupled from hydrodynamic forces for a somewhat arbitrary timescale \citep[e.g.,][]{Vogelsberger+13}.  A wide range of choices is also available for implementing feedback from radiation and jets from accreting supermassive black holes \citep[see][for a review]{SomervilleDave15}. 

Cosmological and idealized simulations, spanning a broad spectrum of physical processes and sub-grid implementations, have been used to study a wide range of CGM properties. Most simulations have been analyzed in different contexts and with different methods. The complementary nature of the two approaches, however, has not previously been exploited. Here, we comparatively analyze in detail a small but representative sample of the existing published simulations. The simulations we selected were not designed with this comparison in mind, so there are certain questions that are beyond the reach of our analysis. However, important physical insight can be gleaned from comparisons of different idealized simulations, different cosmological simulations, and idealized and cosmological simulations.

The overarching goal of the SMAUG\footnote{\href{https://www.simonsfoundation.org/flatiron/center-for-computational-astrophysics/galaxy-formation/smaug/}{Simulating Multiscale Astrophysics to Understand Galaxies}} project is to understand how the key physical processes, which span a huge range of scales, combine to shape the growth of galaxies. As such, the project is in large part focused on understanding the impact of different galaxy formation simulation approaches and the sub-grid models that are used to bridge the vast scales. In the present paper, as part of the first results from SMAUG,\footnote{\url{https://www.simonsfoundation.org/flatiron/center-for-computational-astrophysics/galaxy-formation/smaug/papersplash1}} we advance this goal by comparing in detail the properties of the multiphase CGM---an essential, yet relatively poorly understood aspect of galaxy formation---in simulations that adopt disparate approaches and sub-grid models, and cover a wide range of spatial resolutions. A key result of this comparison for the SMAUG project is that the presence of significant cold gas in the CGM---the most readily observable phase of the CGM and a major source of fuel for future star formation---is dramatically different in the simulations that we analyze below. This points to the need to thoroughly resolve the CGM and accurately include the processes that shape the CGM.

In \autoref{sec:simulations} we introduce the sample of simulations that includes seven idealized simulations and two cosmological simulations. The results of our comparative analysis are presented in \autoref{sec:results} proceeding from a coarse-grained examination of CGM properties to a more granular view. We discuss the implications and context of our findings in \autoref{sec:discussion} and summarize in \autoref{sec:conclusion}.


\section{Simulation Sample} \label{sec:simulations}

\begin{deluxetable*}{lccccccccc}
\tablecolumns{10}
\tablecaption{ Summary of simulations. \label{table:sims}}
\tablehead{
\colhead{Simulation}  & \colhead{Type} & \colhead{$M_{200c}$} & \colhead{$r_{200c}{}^a$}  & \colhead{$T_{200c}{}^b$} & \colhead{$v_{200c}{}^c$} & \colhead{Feedback} & \colhead{$\langle {\rm SFR} \rangle$} & \colhead{$\langle dx \rangle$ in,out$^d$} & \colhead{$\langle dm \rangle$ in,out$^d$} \\
\colhead{label} & \colhead{} & \colhead{$[10^{12} \, M_\odot]$} & \colhead{[kpc]} & \colhead{[$10^{5}$ K]}   & \colhead{[km/s]} & \colhead{Model} & \colhead{$[M_\odot {\rm /yr}]$} & \colhead{[kpc]} & \colhead{$[10^{3} M_\odot]$}}
\startdata
TNG SF          & cosmo.$^*$ & $0.77^{+0.06}_{-0.08}$ & $193^{+5}_{-7}$ & $6.5^{+0.3}_{-0.4}$ & $131^{+3}_{-4}$ &  SF + AGN decoupled wind               &  $2.4^{+1.1}_{-1.3}$   & 1, 10         & 1400, 1400 \\
TNG Q           & cosmo.$^*$ & $0.78^{+0.07}_{-0.05}$ & $194^{+6}_{-4}$ & $6.5^{+0.4}_{-0.3}$ & $132^{+4}_{-2}$ &  SF + AGN decoupled wind               &  $0^{+0.004}_{-0}$   & 1, 10         & 1400, 1400 \\
J12 & cosmo. & 1.42 & 237 & 9.72 & 165 & SF + thermal SN       & ${\sim}5$ & 0.4,2 & 0.4, 1  \\
F17 high $\eta$ & ideal. & $0.71$     & 185  & $6.3$ & 128 & $\eta_m = 2 \, v_{\rm wind} = 700$ km/s  &  $1.50^{+0.28}_{-0.29}$ & 1.4, 1.4 & $15$, $1$ \\
F17 low $\eta$  & ideal. & $0.71$     & 185  & $6.3$ & 128 & $\eta_m =0.3\, v_{\rm wind} = 1200$ km/s &  $1.21^{+0.44}_{-0.36}$            & 1.4, 1.4 & $15$, $1$ \\
L20a SFR3       & ideal. & $1.18$     & 223  & $8.6$ & 151 &    $\eta_m=$1, $\eta_E=$0.3,  $\eta_Z=$0.5                                   &      3         &  0.39, 3.12        &  1.5, 1.5\\
L20b SFR10      & ideal. & $1.18$     & 223  & $8.6$ & 151 &                  $\eta_m=$0.2, $\eta_E=$0.3,  $\eta_Z=$0.5                      &       10        &    0.39, 3.12      & 1.5, 1.5 \\
S20 FIRE & ideal.$^*$ &  1.51 &  242    & 10.1 & 164 &  FIRE & $5.52^{+1.34}_{-0.88}$ &   0.3, 2.8       & 8, 8 \\
S20 Therm & ideal.$^*$ & 1.52 & 243 & 10.2 & 165 & FIRE+Constant $\dot{E}_{\rm thermal}$ & $2.60^{+0.38}_{-0.31}$ &  0.4, 2.7 & 8, 8\\
S20 Turb   & ideal.$^*$ &  1.50 &  242    & 10.1 & 164&   FIRE+Turbulent Stirring &  $2.23^{+0.32}_{-0.19}$ & 0.3, 2.7 & 8, 8 
\enddata
\hspace{6pt}$^*$ Include magnetic fields. \\
${}^{a-c}$Defined in Equations \ref{eq:r200c}--\ref{eq:v200c}.\\ ${}^{d}$The $\langle dx \rangle$ in,out and $\langle dm \rangle$ in,out columns show the average spatial and mass resolution at $0.1 \rvir$ and $\rvir$, respectively. \\
The $\pm$ represents the one sigma variation over time for the idealized simulations and over the population for the TNG SF and Q samples. 
\vspace{-0.3cm}\label{table}
\end{deluxetable*}

In this section we describe the simulations that we compare in this paper. We begin with a general summary followed by a description of specific details for each simulation in the subsequent subsections. For further details readers are encouraged to read the published works from which these simulations are drawn. The most salient properties of the simulations are listed in \autoref{table}.

Broadly, all of the simulations we analyze include the hydrodynamic evolution of the CGM under the influence of the dark matter gravitational field and radiative cooling that includes the impact of the metagalactic UV background. The simulations all include some form of galactic winds that are driven by stellar feedback and/or AGN feedback. The cosmological simulations self consistently include satellite galaxies and nonspherical accretion from the IGM, while the idealized simulations do not. None of the simulations include thermal conduction or cosmic rays.

In all simulations we define the CGM to be comprised of the gas between 0.1 and 1 $\rvir$, where $\rvir$ is a proxy for the virial radius and is defined below in \autoref{eq:r200c}. We analyze the final $z=0$ snapshot of the cosmological simulations. The idealized simulations are analyzed after being averaged over roughly a dynamical time at a sufficiently late time to ensure that the initial transients have diminished. 

\subsection{Cosmological Simulations}

We analyze two cosmological simulations, chosen for their different code types (the moving-mesh code AREPO and the Eulerian grid adaptive mesh refinement code Enzo) and different feedback schemes.

\subsubsection{IllustrisTNG}

The IllustrisTNG simulation suite, and in particular the TNG100 simulation \citep{Marinacci18,Naiman18,Nelson18,Pillepich18,Springel18}, uses the moving-mesh code \textsc{Arepo} \citep{Springel10,Weinberger19} to evolve a box that is $\approx 111 \; \rm{Mpc}$ on each side to $z=0$ from cosmological initial conditions, with a mass resolution of $1.4\times10^6 \; \rm{M_{\odot}}$ per gas cell. For the halos we study in this paper, the typical spatial resolution in the CGM ranges from $\approx 1 \; \rm{kpc}$ near the galaxy to $\approx 10 \; \rm{kpc}$ near the virial radius, and the mass loading factor ($\eta_{\rm M} = \dot{M}_{\rm outflow} / \dot{M}_\star$) of the winds is $\approx 1$ \citep{Pillepich18a,Nelson+19}. The TNG physics model includes star-formation and stellar feedback, black hole formation and AGN feedback, metagalactic UV background \citep{CAFG+09}, and magnetic fields. For more detailed information on the physics model, see \cite{Weinberger17} and \cite{Pillepich18a}.

From TNG100, we use halos defined by the friends-of-friends algorithm \citep{Davis85} and remove all gas bound to satellite subhalos as calculated by the SUBFIND algorithm \citep{Springel01}. All gas velocities are calculated in the center-of-mass reference frame of each central galaxy (defined as all stars in a halo's central subhalo) using proper coordinates, thus removing effects of both cosmological expansion and motion in the cosmic web, neither of which occurs in the idealized simulations.

We select 50 star forming and 50 quiescent halos with total halo masses of $\approx 10^{12} \; \rm{M_{\odot}}$ at $z=0$. We define halos as being star-forming (quiescent) if their central galaxy has a specific star-formation rate above (below) $10^{-11} \; \rm{yr^{-1}}$ at $z=0$. The resulting star-forming sample has a median log sSFR$ = -9.95^{+0.17}_{-0.23}$, and the median of the quiescent sample has no current star formation, with the 84th percentile extending up to $10^{-12.9} \; \rm{yr^{-1}}$. We refer to the star forming and quiescent samples as TNG SF and TNG Q, respectively.

\subsubsection{Joung et al. 2012 simulation}

\cite{Joung+12} \citepalias[hereafter][]{Joung+12} carried out a high-resolution zoom simulation of a single halo with mass $\Mhalo = 1.42 \times 10^{12}$ $\Msun$ identified from 25 $h^{-1}$ Mpc comoving volume using parameters consistent with WMAP5 cosmological parameters \citep{Komatsu+09}.  Other results from this simulations were also presented in \citet{Fernandez+12}. The halo was modeled using the Enzo cosmological hydrodynamics code \citep{bryan14} and employed adaptive mesh refinement to obtain high resolution in the zoom region. This simulation employed a $128^3$ root grid with four additional (static) levels covering the initial $\sim (5 h^{-1}$ Mpc)$^3$ Lagrangian volume of the halo, achieving a dark matter particle mass in the refined region of $1.7 \times 10^5$ $\Msun$, and an initial cell (gas) mass of $2.8 \times 10^4$ $\Msun$.  Six additional levels of refinement (for a total of 10) were added such that refinement was triggered whenever a cell contained more than 4 times the initial dark matter or gas masses, down to a best cell size of 270 comoving pc. In the CGM, most cells have a mass of about 1000 $\Msun$ and resolution that goes from 270 pc to 2 kpc at the virial radius.

The simulation included a non-equilibrium chemical network involved ionized states of H and He, as well as a metallicity field which was used to compute metallicity-dependent cooling down to 10 K. A metagalactic UV background \citep{Haardt+96} with local self-shielding, and a diffuse form of photoelectric heating. Sites of star formation were identified as gas with a density larger than $7 \times 10^{-26}$ g cm$^{-3}$ and a mass larger than the local Jeans mass.  To prevent large number of low-mass stellar particles, stars were generated stochastically with a minimum stellar particle of $10^5$ $\Msun$.  Gas was converted into stars with an efficiency per free-fall time of 3\%. Supernova feedback was ejected in the form of thermal energy spread over the 27 local cells weighted inversely by the gas density in that cell; this energy was added over a dynamical time. In addition to energy, metals were added to the gas with a yield of 0.025. Young stars also contributed to a diffuse FUV background that sourced the photoelectric heating.

The results presented here are from a single snapshot of the simulation at $z=0$. As described in more detail elsewhere \citep{Fernandez+12}, at $z=0$ the system is dominated by a cold, rotating gaseous disk with a large stellar component. The central stellar mass is somewhat larger than typically found for halo masses of this size (i.e. it may suffer from an overcooling problem, although it is hard to be sure with only a single sample) and the rotation curve in the central few kpc is higher than observed for most disk systems.  

\subsection{Idealized Simulations}

We use seven idealized simulations coming from three studies that use different feedback models, simulation geometries, included processes, and numerical methods, as described below.

\subsubsection{Fielding et al. 2017 simulations}

\cite{Fielding+17} \citepalias[hereafter][]{Fielding+17} simulated the evolution of the CGM of eight halos ranging from $10^{11}$ to $10^{12}$ $\Msun$ under the combined influence of large scale accretion from the intergalactic medium and galactic winds. In this work we only focus on the two $10^{12}$ $\Msun$ halos. These three-dimensional hydrodynamic simulations were performed using {\tt athena 4.2} \citep{Stone+08,Gardiner+08}. 

The simulations were performed using a Cartesian grid with a box size of 1.44 Mpc. Static mesh refinement was employed to give higher resolution toward the center of the domain. A nested cubes geometry was adopted such that the resolution doubles for each factor of two closer to the center of the domain. The highest resolution region which was 2 $\rvir=360$ kpc on a side had a spatial resolution of $\Delta x = 1.4$ kpc. 

Accretion from the intergalactic medium was included by feeding gas into the halo at the turn around radius $3.5 \rvir$. This additional mass was added in a predominantly spherical manner although density fluctuations were added to break perfect spherical symmetry. The accretion rates were chosen to match the mean rates measured in the Millennium simulation \citep{Millennium} by \citet{McBride+09} by multiplying the dark matter accretion rate by the cosmic baryon fraction $\sim 0.16$. For the two halos we consider here this corresponds to a large scale baryonic accretion rate of 7 $\Msun$/yr.

Galactic winds were modeled in the simulations by measuring the flux of mass through a small sphere of radius 8 kpc $= 0.04 r_{\rm 200c}$ in the center of the domain and then ejecting some fraction of that mass back out into the domain at some predetermined velocity. In this way all of the complicated, and computationally expensive galaxy formation processes were ignored and parameterized entirely by two parameters: the mass loading factor of the wind $\eta_{\rm M}$ and the velocity of the wind $v_{\rm wind}$. The two simulations we analyze in this work had $\eta_{\rm M} = 2$ and $v_{\rm wind} = \sqrt{3} v_{\rm esc} = 700$ km/s, and $\eta_{\rm M}=0.3$ and $v_{\rm wind}= 3 v_{\rm esc} = 1200$ km/s, which we refer to as high $\eta$ and low $\eta$, respectively. These wind velocities correspond to energy loading factors of $\eta_{\rm E} = \dot{E}_{\rm wind} / \dot{E}_{\rm SN} = $ 0.245 and 0.108 for the high and low $\eta$ simulations.

A static NFW gravitational potential was employed to model the impact of dark matter \citep{NFW}. No contribution was added for the central galaxy since the main focus was out in the halo where dark matter dominates. All gas was assumed to have one-third solar metallicity and to be in ionization equilibrium. The cooling (and heating) rates were taken from the ionization equilibrium tables published by \cite{Wiersma+09} who adopted the \cite{HaardtMadau01} metagalatic UV background.

When analyzing these simulations we use the average halo properties between 6 and 9 Gyr. At this late time any imprint of the initial conditions are sufficiently diminished, and by averaging over two dynamical times transient variations are washed out. The median star formation rate during this period is $\sim 1.4 $ $\Msun$/yr for both choices of feedback model.

\subsubsection{Li \& Tonnesen 2020a,b simulations}

\cite{Li+20a} \citepalias[hereafter][]{Li+20a} and Li \& Tonnesen (2020b in preparation; hereafter L20b) ran a suite of simulations studying the CGM in $\Mhalo=1.18\times10^{12}$ $\Msun$ halos with varying initial densities and constant SFRs in the disk. The \citetalias{Li+20a} simulation had a star formation rate of 3 M$_\odot$/yr (hereafter referred to as the \citetalias{Li+20a} SFR3 simulation), and the L20b simulation had a star formation rate of 10 $M_\odot$/yr (hereafter referred to as the L20b SFR10 simulation).

The hydrodynamic equations are solved by the Eulerian code Enzo \citep{bryan14}, using the finite volume piece-wise parabolic method \citep{colella84}. The fiducial box size is 800 kpc on each side. Static mesh refinement was used throughout the simulation. The spatial resolution is progressively higher toward the center of the box, which is 0.39 kpc for the inner (50 kpc)$^3$, 0.78 kpc for the inner (100 kpc)$^3$, and so on.  

Star formation is not modeled directly in these simulations. Instead, outflows are injected as discrete events to a small region near the galaxy disk. The locations of outflows are different for each event, which are randomly selected within $R_{SF}$ in radius. The time intervals between these outflow events are constant, $\Delta t=9.9$ Myr.  For each outflow event, the injected region is two hemispheres with radii of 3 kpc a few kpc above and below the galaxy disk plane. Only hot outflows are added.  The mass, energy, and metal loading factors of the outflows are set as $\eta_{\rm M} = 1.0$, $\eta_{\rm E}=0.3$, $\eta_Z = \dot{Z}_{\rm out} / \dot{Z}_{\rm SN}=0.5$ for the \citetalias{Li+20a} SFR3 simulation and $\eta_{\rm M} =0.2$, $\eta_{\rm E}=0.3$, $\eta_Z=0.5$ for the L20b SFR10 simulation.  The radius in the galaxy within which SF regions may occur is $R_{\rm SF}$ = 8 kpc for \citetalias{Li+20a} SFR3 and 2 kpc for L20b SFR10, indicating that the SF is widespread in the galaxy disk for the former and more centrally concentrated for the latter.

There is a static gravitational potential. The potential includes a dark matter (DM) halo, a stellar disk, and a stellar bulge. The parameters of the potential follow those of the Milky Way (MW). The DM halo is assumed to have a \citet{burkert95} profile. The mass distribution of the stellar disk has a Plummer-Kuzmin functional form \citep{MiyamotoNagai75}, and the bulge is modeled as a spherical \citep{Hernquist,hernquist93} profile.

The initial gas in the simulation box only includes a hot halo component, and there is no cool gaseous disk within the galaxy, as the focus is the circumgalactic medium. The halo gas has a uniform temperature of $10^6$ K, similar to the virial temperature of the DM halo, and a uniform low metallicity 0.2 Z$_\odot$. Gas density is set to be in hydrostatic equilibrium with the DM halo potential, with an inner cutoff at $R=40$ kpc. Gas inside of this radius has a uniform density equal to that at $R=40$ kpc. Inflows are not included, and the normalization of the mass of the pre-existing hot halo is chosen such that, after the CGM reaches a steady state, the X-ray luminosity matches that of the observations for galaxies with similar masses and SFRs.

The Grackle library \citep{smith11} was used to calculate the cooling rate of the gas, assuming ionization equilibrium with the metagalactic UV background of \cite{haardt12}. The cooling is metallicity-dependent, with the outflows having a metallicity of $1.4 Z_\odot$. The simulation outputs used in this paper are averaged $t=3$ -- 5 Gyr for the SFR3 run and over 0.7- 1.5 Gyr for the SFR10 run. 

\subsubsection{Su et al. 2020 simulations}

\cite{Su+20} \citepalias[hereafter][]{Su+20} studied the cooling flow properties for galaxy of $10^{12}-10^{14} M_\odot$ with various AGN feedback toy models on top of FIRE-2 stellar feedback model. Here we focus on three $\Mhalo = 1.5 \times 10^{12} M_\odot$ halo simulations. The details of the simulations `S20 FIRE', `S20 Therm', and `S20 Turb' are described in \citetalias{Su+20} respectively as the `m12-Default', `m12-Th-core-43-wide', and `m12-Turb-core-wide'. 

The simulations use {\sc GIZMO}, in its meshless finite mass (MFM) mode \citep{2015MNRAS.450...53H}. The mass resolution is 8000 $M_\odot$, and the average spatial resolution is $\sim 0.3\, {\rm kpc}$ at $0.1 \,r_{200c}$ and $\sim 3 \,{\rm kpc}$ at $r_{200c}$.

All three \citetalias{Su+20} simulations use the FIRE-2 implementation of the Feedback In Realistic Environments (FIRE) physical treatments of the ISM and stellar feedback, the details of which are given in \citet{FIRE2,Hopkins+18} along with extensive numerical tests.  Cooling is followed from $10-10^{10}$K, including the effects of photoionization heating using the \citet{CAFG+09} metagalactic UV background model. The stellar feedback model includes: (i) radiative feedback, (ii) stellar winds, (iii) and Type II and Ia SNe. 

In the \citetalias{Su+20} Therm simulation, in addition to the FIRE model, a constant heating rate per unit mass is added following a spherically-symmetric Gaussian distribution (centered on the BH at the galaxy center) with a scale length of 14 kpc. The total energy injection in this region is $\sim 2\times10^{43}{\, \rm erg\, s}^{-1}$.

In the \citetalias{Su+20} Turb simulation, solenoidal turbulence was driven directly following the ``turbulent box'' simulations in \citet{2012MNRAS.423.2558B}. Turbulence is driven in Fourier space as an Ornstein-Uhlenbeck process \citep[see][] {2009A&A...494..127S,2010A&A...512A..81F,2010MNRAS.406.1659P} with characteristic driving wavelength ($\lambda=2\pi/k$) set to $1/2$ of the halo scale radius. The driving varies radially following is a Gaussian function with a scale length of $\sim$ 40kpc. The total energy input is $\sim10^{40}{\rm erg\, s}^{-1}$

Initially the DM halo, stellar bulge, and stellar disc are set following \cite{1999MNRAS.307..162S}, assuming a spherical NFW \citep{NFW} profile dark matter halo with a scale length of 20.4 kpc, and a stellar bulge following a \cite{Hernquist} profile with a scale length of 1 kpc. 
Exponential, rotation-supported gas and stellar disks were initialized with scale lengths of 6 kpc and 3 kpc, respectively, and scale-height 0.3 kpc for both. The gas temperatures were initialized to pressure equilibrium \citep{2000MNRAS.312..859S}, and an extended spherical, hydrostatic gas halo with a $\beta$-profile (of scale-radius 20.4 kpc, $\beta=1/2$) and rotation at twice the net DM spin (so $\sim 10-15\%$ of the support against gravity comes from rotation, with the rest provided by thermal pressure resulting from the virial shock). The initial metallicity drops from solar ($Z=0.02$) to $Z=0.001$ with radius as $Z=0.02\,(0.05+0.95/(1+(r/20\,{\rm kpc})^{1.5}))$. The initial magnetic fields are azimuthal with $|{\bf B}|=0.3\,\mu{\rm G}/(1+(r/20\,{\rm kpc})^{0.375})$ (extending throughout the CGM).

All the \citetalias{Su+20} simulations are run for 2.5 Gyr (longer than the original \cite{Su+20} runs for stability at very large radius), and the results below are the averaged values over 2.4-2.45 Gyr. 

\section{Results}\label{sec:results}

\begin{figure*}
\centering
\includegraphics[width=0.9\textwidth]{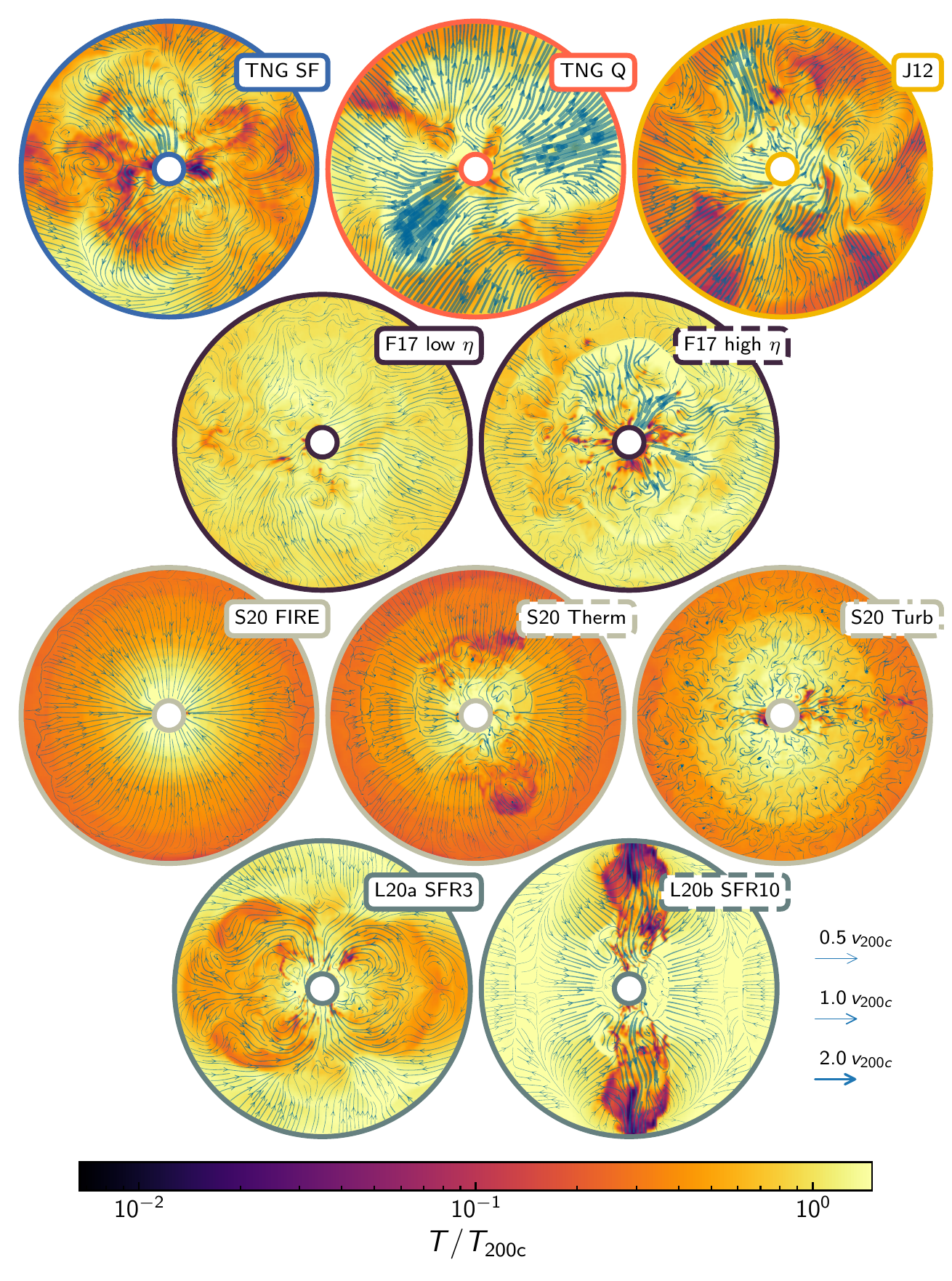}   
\caption{CGM temperature relative to $\Tvir$ from 0.1 to 1 $\rvir$. Velocity is traced by the blue streamlines that vary in thickness proportional to the velocity magnitude relative to $\vvir$. A single example is shown for the TNG SF and Q samples. The halos are oriented perpendicularly to outflow/minor axis (except for the \citetalias{Fielding+17} halos that have no preferred axis). The cosmological simulations exhibit a wide range of temperatures throughout their halos and have large scale velocity asymmetries. The presence of cold gas in the idealized simulations is less wide spread and closely tied to feedback from the central galaxy.}
\label{fig:all_maps}
\end{figure*}

In this section we present the comparative analysis of the seven idealized simulations (\citetalias{Fielding+17} high $\eta$ and low $\eta$, \citetalias{Li+20a} SFR3 and L20b SFR10, and \citetalias{Su+20} FIRE, Therm, and Turb), the AMR cosmological zoom simulation \citepalias{Joung+12}, and the one hundred TNG halos split into 50 star forming (TNG SF) and 50 quiescent (TNG Q). We begin by defining the normalizations we will use throughout and presenting exemplary maps of the CGM temperature and velocity fields to provide an intuitive basis for the quantitative analysis that follows. The next subsection (\S\ref{sec:median_profiles}) focuses on a comparative examination of the median CGM profiles. In the second subsection (\S\ref{sec:temperature_distribution}), we address the temperature distribution as a function of radius. This leads us, in the third subsection (\S\ref{sec:1D_phase}), into a deeper study of the mass distribution as a function of temperature $T$, radial velocity $v_r$, and pressure $P$ in the inner and outer halo. We end our analysis in \S\ref{sec:2D_phase} with a look at the joint distribution of pressure $P$ and an entropy-like quantity (which we refer to in this paper as entropy) $K=P n^{-5/3} = \kb T n^{-2/3}$ in the outer halo.

In most of the following analysis we normalize quantities by their approximate virial values \citep{Kaiser86} using the ``200c'' definitions, which minimizes the apparent differences of various quantities due to minor differences in halo mass across simulations. Radial coordinates are normalized by 
\begin{align}
\rvir =& \left( \frac{G \Mhalo}{100 H_0^2} \right)^{1/3}  = 211 \,\, \rm{kpc} \,\, M_{12}^{1/3} \label{eq:r200c}
\end{align}
where $\Mhalo = M_{12}10^{12} \Msun$, and $H_0 = 67.4$ km/s/Mpc \citep{Planck18}. Temperature is normalized by 
\begin{align}	
\Tvir =& \mump G \Mhalo / (2 \kb \rvir ) 	\\ 
=& 7.7\times10^5~{\rm K}~M_{12}^{2/3}. \nonumber 
\end{align}
Velocities are normalized by 
\begin{align}
\vvir =& \left( \frac{G \Mhalo}{\rvir} \right)^{1/2} = 143 \, M_{12}^{1/3} \, {\rm km/s}. \label{eq:v200c}
\end{align}
Number density is normalized by $\nvir = \fb 200 \rho_c / \mump = 2.8 \times 10^{-4} \, {\rm cm}^{-3}$. Pressure is normalized by $\Pvir = \nvir \kb \Tvir = 216~M_{12}^{2/3}~\kb~{\rm K}~{\rm cm}^{-3}$, and entropy is normalized by $\Kvir = \nvir^{-2/3} \kb \Tvir = 1.8\times10^{8}~M_{12}^{2/3}~\kb~{\rm K}~{\rm cm}^{2}$. The velocity asymmetries and turbulent velocities are characterized by the velocity dispersion, which is calculated relative to the net average velocity. To be precise,
\begin{align}
\sigma_v^2 =\; & \langle (v_r - \langle v_r \rangle_M)^2 \rangle_M + \langle (v_\theta - \langle v_\theta \rangle_M)^2 \rangle_M  \nonumber \\ 
&+ \langle (v_\phi - \langle v_\phi \rangle_M)^2 \rangle_M 
\label{eq:vrms}\end{align}
where $\langle \cdot \rangle_M$ denotes a mass weighted average over a spherical shell.

\autoref{fig:all_maps} illustrates the temperature and velocity fields in the CGM of the simulations. The maps show the mass-weighted temperature from 0.1 to 1 $\rvir$ in a projection 0.1 $\rvir$ thick centered on the galaxy. The blue streamlines trace the velocity with the thickness proportional to the velocity magnitude divided by $\vvir$. Rather than show all 100 TNG halos we pick one representative example from the TNG SF and Q samples. 

Relative to the idealized simulations, the cosmological simulations have gas at a wide range of temperatures distributed throughout their halos. The CGM in the idealized simulations, on the other hand, is mostly filled with hot, virial temperature gas. The cold phase in the idealized simulations is much less wide spread and is closely tied to galactic feedback. The \citetalias{Su+20} simulations exhibit a steeper decline in the temperature of the hot phase relative to the other idealized simulations, which is in large part due to the more concentrated mass profile and initial conditions. The velocity field in the idealized simulations is primarily comprised of small scale turbulence and general spherical symmetry whereas the cosmological simulations exhibit relatively smooth flows on small scales and large scale velocity asymmetries. These (and other) trends and differences are further clarified in the analysis presented in the subsequent sections.


\begin{figure*}
\centering
\includegraphics[width=\textwidth]{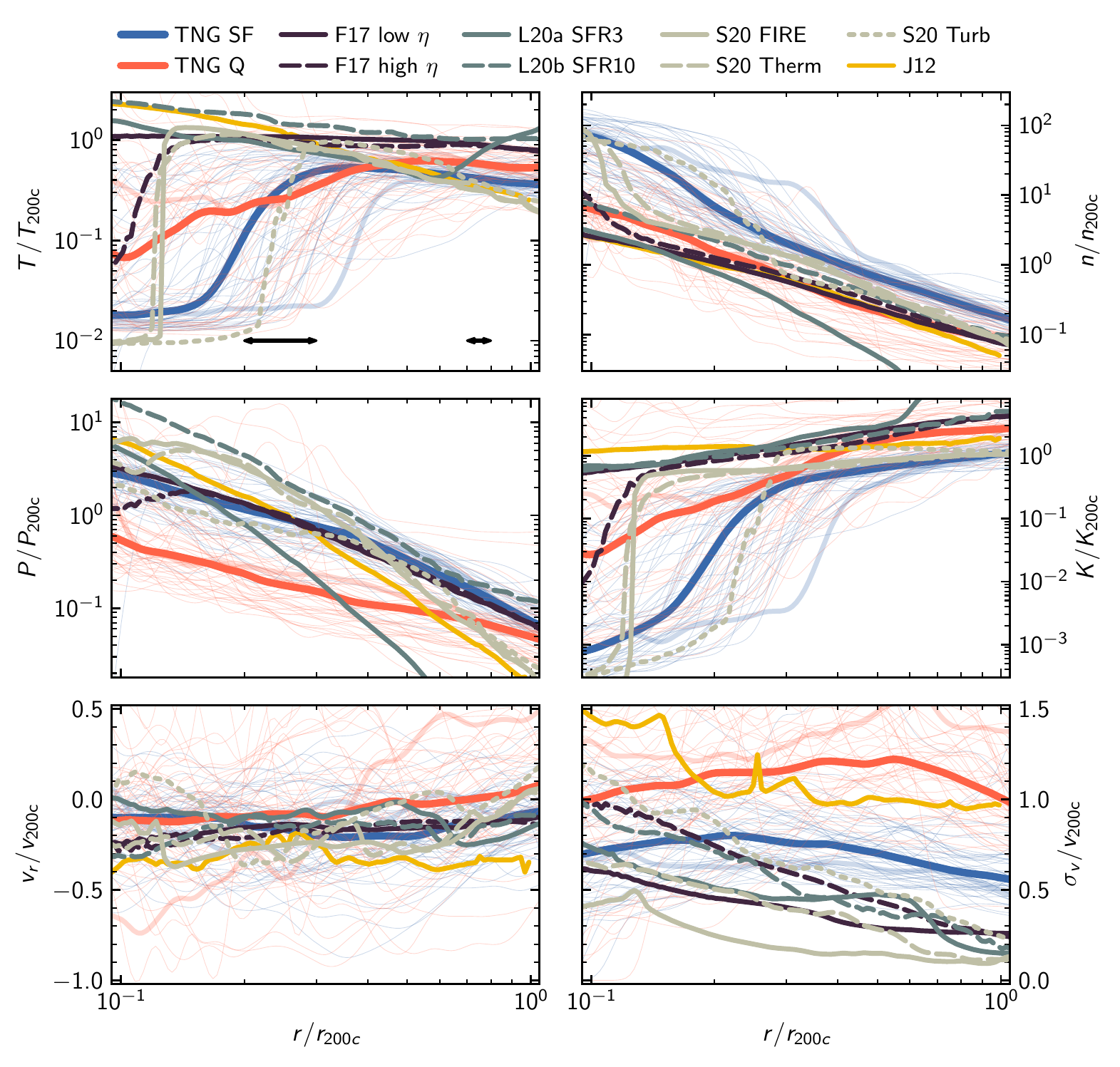}   
\caption{From left to right and top to bottom, the virial normalized mass-weighted median temperature, number density, pressure, entropy, radial velocity, and velocity dispersion profiles are shown. The thin blue (orange) lines show the individual TNG SF (Q) halos, with the thicker semi-transparent line tracing the individual example shown in Figures \ref{fig:all_maps}, \ref{fig:all_phase}, and \ref{fig:pressure_entropy}. The thick opaque blue (orange) lines show the median of TNG SF (Q) halos. The \citetalias{Fielding+17} low and high $\eta$ simulations are shown in dark purple solid and dashed lines, respectively. The \citetalias{Li+20a} SFR3 and L20b SFR10 simulations are shown in medium blue-gray in solid and dashed lines, respectively. The FIRE, Therm, and Turb feedback \citetalias{Su+20} simulations are shown in light yelowish-gray in solid, dashed, and dotted lines, respectively. The \citetalias{Joung+12} cosmological zoom-in simulation is shown in yellow. In simulations with angular momentum, rotation becomes important at radii $<0.2 \rvir$ so we focus mostly on larger radii. The arrows on the bottom of the top left panel denote 0.2-0.3 $\rvir$ and 0.7-0.8 $\rvir$, which are the shells that we look at in the subsequent figures. These profiles highlight some of the key differences and similarities between the bulk structure of the halos of these markedly different simulations. The profiles fall somewhere between isentropic and isothermal through much of their volume with roughly the same density normalization. Among the most striking differences is that the quiescent TNG simulations are significantly under pressurized relative to the other halos, and that the cosmological simulations have significantly larger velocity dispersions than any of the idealized simulations, particularly in the outer halo.}
\label{fig:all_profiles}
\end{figure*}

\subsection{Median Radial Profiles}\label{sec:median_profiles}

\autoref{fig:all_profiles} shows the radial dependence of the virial-normalized mass-weighted median temperature, number density, pressure, entropy, radial velocity, and velocity dispersion of the halo gas in all of the simulations from 0.1 to 1 $\rvir$. 
We emphasize that the CGM in these simulations is strongly multiphase, even at a given radius (as seen for the temperature in Figure \ref{fig:all_maps}), thus their distributions are wide, which is not reflected in the median profiles shown here. We will examine full distributions for some of these quantities later in the paper. 

\subsubsection{Temperature profiles}
Starting with the temperature in the top left panel a few trends are immediately clear. First and most apparent, some of the simulations show a marked decrease in temperature in the inner halo which indicates that cold gas is beginning to dominate the mass budget. This mostly occurs in the simulations with rotation (TNG, \citetalias{Joung+12}, and \citetalias{Su+20}) and is likely a result of reaching the outermost edges of the disk where angular momentum support begins to become appreciable and can therefore hold the cold gas up against gravity. This indicates that the CGM phase structure out to $r\lesssim 0.2 \rvir$ is impacted by angular momentum. Although the TNG Q sample contains more halos with a hot inner CGM than the TNG SF sample, which leads to a higher median temperature in the inner regions, both populations have systems in which the $10^4$ K phase dominates far out into the halo and systems in which the virialized component dominates down to below $0.1\,\rvir$. Interestingly, the \citetalias{Joung+12} cosmological zoom simulation remains hot all the way down to the innermost portion of the halo. This difference relative to most of the TNG halos is likely a result of the purely thermal feedback implemented in \citetalias{Joung+12} as opposed to the hybrid thermal-kinetic feedback in TNG. 

Shifting our attention farther out in the halo where the bulk of the CGM resides, we find that the median CGM temperature of the different simulations are within a factor of $\sim 2-3$ of $\Tvir$. Given the diversity of these simulations this agreement is non-trivial. Moreover, the median halo temperatures all decrease slowly with radius, falling by a factor of $\sim 2$ from 0.2 to 1~$\rvir$. This is a result of the halo gas temperature roughly following the circular velocity of the halo, as is expected for a virialized halo. Accordingly, the (relatively minor) differences in temperature profiles are likely due to the differences in the dark matter profiles and the associated circular velocity profiles. The temperature profiles of the \citetalias{Fielding+17} simulations decrease more slowly than the other idealized simulations because these halos have a more extended dark matter halo. The \citetalias{Li+20a} SFR3 simulation has an uptick in temperature at 0.6 $\rvir$ because the outflows from the central galaxy never reach that far, so this gas is left over from the initial conditions. 

The median of the TNG SF and Q samples are systematically lower than the idealized simulations. Part of this relative decrement is, as we show below, due to the presence of more cold gas at large radii in the TNG halos that brings the median down. The median temperature of the Q sample is somewhat higher than the median temperature of the SF sample. However, the TNG Q sample exhibits significantly more variation from halo to halo than the TNG SF sample, as shown by the thin lines.

\subsubsection{Density profiles}
The median density profiles of the halo gas in all the simulations, shown in top right panel of \autoref{fig:all_profiles}, also exhibit a strong similarity in shape and normalization. This is evidence that the CGM of all these halos contain about the same fraction of their host halos' mass. The exception to this similarity is the star forming TNG halos that have roughly two to three times higher density throughout the bulk of the halo than any of the idealized or quenched TNG halos (particularly in the inner CGM). As with the temperature, the density of the quenched TNG halos exhibits more variability from halo to halo than the SF population, with many halos having CGM densities an order of magnitude lower than the least dense halo from the SF sample. Note that the \citetalias{Li+20a} SFR3 simulation has very low density beyond $\sim 0.4 \rvir$ because the winds, due to their low specific energy, cannot reach out that far to build up an appreciable halo, so all that exists is the low density (high entropy) gas from the initial conditions. 

The median halo density throughout the bulk of the halos falls off as roughly $r^{-3/2}$, which is consistent with Milky Way X-ray observation constraints \citep{Miller+15}, as well as analytic models \citep[e.g.,][]{Faerman+17,Stern+19}.

\subsubsection{Pressure and Entropy profiles}
The pressure and entropy profiles---shown in the middle panels of \autoref{fig:all_profiles}---give a clearer view of the differences and similarities between the simulated halo gas properties. As we discuss below in reference to \autoref{fig:pressure_entropy}, working with pressure and entropy is particularly useful because, in principle, radiative cooling should remove entropy but leave the pressure unchanged (in the limit where cooling is fully resolved). 

The median pressures of the TNG SF sample, the \citetalias{Joung+12}, and the idealized simulations agree with each other much more than the median TNG Q sample, which is significantly lower and shallower. The pressure of the TNG SF sample, the \citetalias{Joung+12}, and the idealized simulations falls off with radius as $r^{-2}$ to $r^{-3/2}$ (with the exception of the \citetalias{Li+20a} SFR3 simulation which falls off even more steeply). 

The importance of feedback in setting the CGM properties is reflected in the differences in pressure profiles. The \citetalias{Joung+12} simulation has an appreciably steeper pressure profile than the TNG SF sample. And, even more apparent, the median pressure of the TNG Q sample has the lowest central value and the shallowest slope, falling off roughly linearly with radius. Around $\rvir$ the TNG SF and Q samples converge to roughly the same value. Strong AGN feedback has been shown to be the dominant cause of quenching star formation in TNG. It is therefore likely that the significant under-pressurization of the halo gas in the TNG Q sample is a result of the AGN feedback ejecting material from the inner halo. This is consistent with past analysis of the TNG simulations \citep{Davies+19}. The material that does remain in the CGM of the TNG Q halos is likely supported in large part by an effective turbulent support, which is consistent with the large velocity dispersions in these halos, as shown in the bottom right panel of \autoref{fig:all_profiles}.

The entropy profiles in the outer halo of all the simulations increase approximately linearly with radius. This roughly linear entropy profile is expected for a subsonic cooling flow \citep{Stern+19}, precipitation-limited models \citep{Voit2019}, or from cosmological assembly \citep{TozziNorman2001,Voit+05}. The entropy at large radii in the \citetalias{Li+20a} (which reflects the initial conditions), \citetalias{Fielding+17}, \citetalias{Joung+12}, and the TNG Q sample are a factor of $3-10$ higher than in the outskirts of the \citetalias{Su+20} and TNG SF simulations. The \citetalias{Su+20} entropy profile at large radii is quite similar to the initial conditions, but the difference in the entropy at large radii between the TNG SF and Q samples is again likely due to the AGN feedback driven expulsion of gas in the TNG Q sample. On the other hand, the relatively high entropy in the idealized simulations likely reflects the lack of cooling and multiphase gas at large radii, as we will discuss below. This is supported by the relatively scarce cool multiphase gas at large radii seen in the idealized simulations in \autoref{fig:all_maps}. The slight difference in the normalization and shape of the entropy profiles at large radii may also be due to slight differences in the circular velocity profile.

\subsubsection{Velocity profiles}
The median radial velocity of the gas in all halos but the TNG Q sample exhibits a slight inflow throughout the full volume of the halos, as shown in the bottom left panel of \autoref{fig:all_profiles}. This reflects a net transfer for material from the CGM (and, presumably, beyond from the IGM) to the ISM. This inflow provides the fuel for ongoing star formation and AGN activity. The two TNG samples exhibit strong variation from halo to halo, especially in the Q sample. This variability may be accentuated by the fact that the idealized simulations are averaged over a significant time window whereas the cosmological simulations are taken at a single time. The \citetalias{Joung+12} simulation is consistent with some of the more inflowing TNG simulations and is generally more inflowing than all of the idealized simulations, especially near $\rvir$.

\begin{figure*}
\centering
\includegraphics[width=\textwidth]{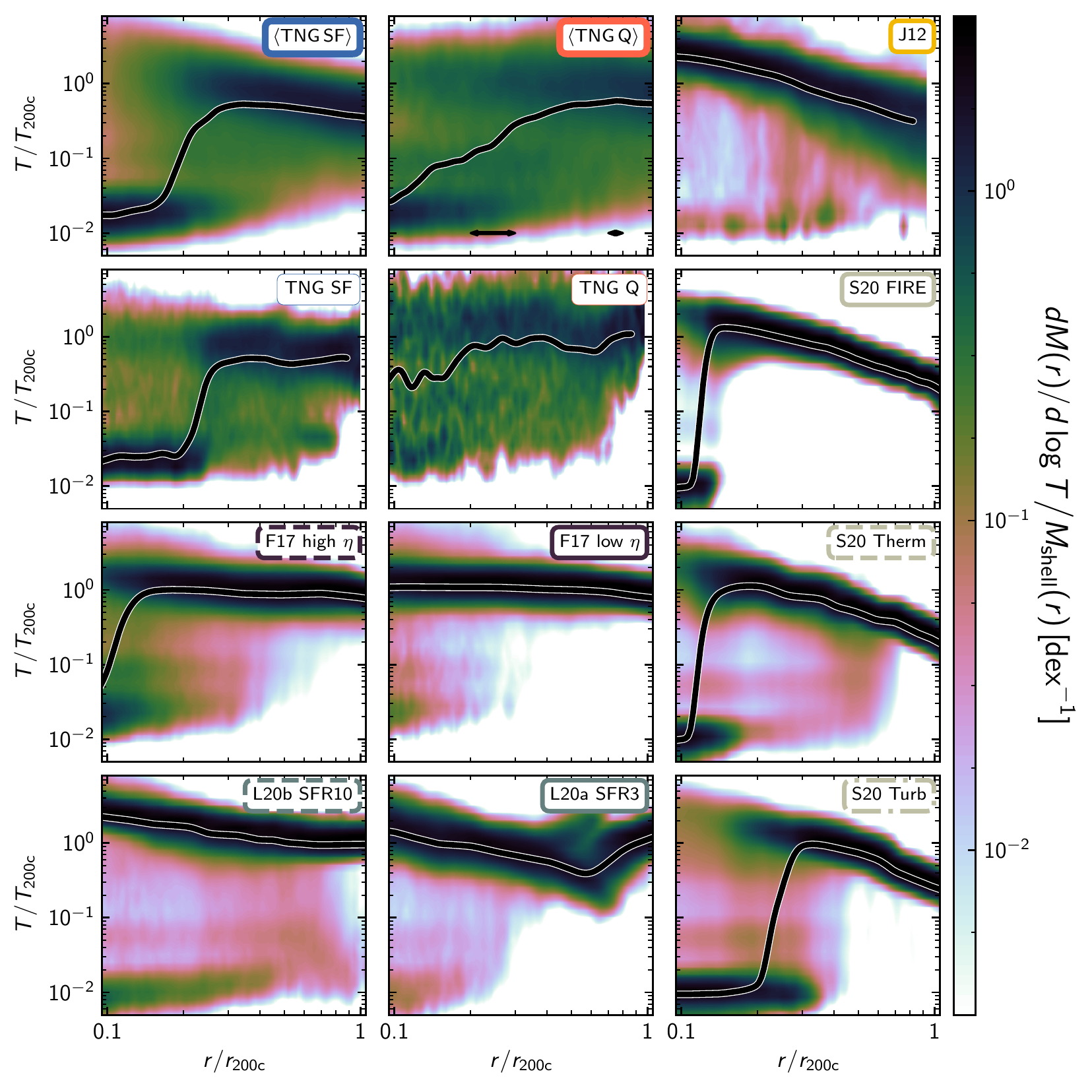}   
\caption{The colored histograms show the temperature-radius distributions. The distribution at each radius is normalized to the amount of mass in that radial shell to highlight the relative shape of the distributions as a function of radius. The thick outlined line in each plot shows the same median profiles shown in \autoref{fig:all_profiles}. The line around the plot label corresponds to the line style used in all other plots. The median of the TNG SF and Q halos are shown in the top row, and individual examples are shown in the second row (same example halos as in other figures). Beyond 0.2 $\rvir$ the idealized simulations are mostly single phase with a narrow distribution around the virial temperature and with only a small cold phase, with the exception being the strongest feedback cases. On the other hand the TNG simulations all have broad temperature distributions spanning up to two orders of magnitude in temperature at all radii regardless of star formation rate. In the inner halo the \citetalias{Joung+12} simulation is similar to the idealized simulations with a small cold phase, but near $\rvir$ the temperature distribution broadens significantly and resembles the TNG simulations with significant cold and hot gas. The arrows on the bottom of the top middle panel denote 0.2-0.3 $\rvir$ and 0.7-0.8 $\rvir$, which are the shells that we look at in the subsequent figures.}
\label{fig:all_phase}
\end{figure*}

The radial profile of CGM velocity dispersion $\sigma_v$, shown in the bottom right panel of \autoref{fig:all_profiles}, differs strongly between the different simulations. The velocity dispersion, which is calculated using \autoref{eq:vrms}, encapsulates both turbulent motions and large scale velocity asymmetries. As shown in \autoref{fig:all_maps}, the velocity field in the cosmological simulations tends to exhibit more large scale velocity asymmetries, whereas the velocity dispersion in the idealized simulations tends to arise more from smaller scale random motions. In the idealized simulations feedback is the only mechanism that can lead to departures from symmetry and thereby increase the velocity dispersion. In the cosmological simulations, however, the motion of satellite galaxies and asymmetrical/filamentary accretion from the IGM naturally produces significant velocity dispersions in addition to the feedback induced stirring.

In the cores of all of the halos---cosmological and idealized---there is an appreciable velocity dispersion on the order of $\vvir$. This large central velocity dispersion reflects the ability of feedback to efficiently stir the inner halo gas regardless of the presence of other processes. Beyond 0.2 $\rvir$, however, the velocity dispersion in the cosmological simulations is larger than any of the idealized simulations, which points to the increasing importance of other processes farther out in the halo. 

The central velocity dispersion reaches as high as a few $\vvir$ in many of the TNG Q halos and the \citetalias{Joung+12} simulation.  The velocity dispersion in the \citetalias{Joung+12} halo and the TNG Q  halos are $\gtrsim 2 \times$ larger than in the TNG SF sample with velocity dispersions remaining on the order of $\vvir$ out to $\sim \rvir$. With increasing halo radius the velocity dispersion in idealized simulations drops much faster than in any of the cosmological simulation halos, reaching values of $\sim 0.1 \vvir$ at $\rvir$. Beyond $\sim 0.4 \rvir$ the TNG SF halos have turbulent velocities at least twice as large as those in the idealized simulations. This dramatic difference in the velocity dispersions of the \emph{outer} halo gas in the idealized and cosmological simulations is reflected in many of the distributions we look at below and points to processes beyond galactic feedback that are important for setting the \emph{outer} CGM structure, such as IGM accretion and halo substructure.

\subsection{Radial Temperature Distribution}\label{sec:temperature_distribution}

Having examined the median profiles, we next turn to the mass-weighted radius-temperature distribution of the simulations. \autoref{fig:all_phase} shows this for the seven idealized simulations, the \citetalias{Joung+12} cosmological zoom-in simulation, two individual TNG halos, and the median of the TNG SF and Q samples. In each radial shell, 0.01 $\rvir$ wide, the fraction of the shell mass per temperature bin is shown with the colored histogram. The thick line shows the same median temperature profile shown in \autoref{fig:all_profiles}. 

In all of the simulations, the majority of the mass beyond $\sim 0.2 \rvir$ resides in a virialized component. In the idealized simulations the $T\sim\Tvir$ component mass fraction dominates the amount contained in lower temperature phases by at least an order of magnitude at all radii (except in the very center of the \citetalias{Su+20} simulations where rotational support begins to replace thermal pressure support). The picture is very different in the TNG halos---at all radii there is a significant sub-virial component which extends down to $\sim 10^{-2}~\Tvir \sim 10^4$ K. Although the virial component dominates the total mass beyond $\sim 0.2 \rvir$, the fraction contained in the cold, sub-virial component is orders of magnitude more than in the idealized simulations and extends all the way out to $\rvir$. The presence of the significant cold phase at large radii draws the mass-weighted median down in the TNG samples, which means the hot, virial component, which fills most of the volume, in the idealized simulations and TNG halos are more similar than indicated by the profiles shown in the top left panel of \autoref{fig:all_profiles}. 

The temperature distribution in the halo of the \citetalias{Joung+12} simulation is dominated by the hot phase in the inner halo, similar to what is seen in the idealized simulations. In the outer halo, however, unlike in the idealized simulations the temperature distribution broadens significantly. By $\sim \rvir$ the \citetalias{Joung+12} simulation has a cold phase similar to the TNG simulations. \emph{This supports the statement that the phase structure in the inner halo is predominantly set by feedback processes while at large radii other processes, which are cosmological in nature and not included in the idealized simulations, take precedence. }

Somewhat counter-intuitively, the CGM around the TNG Q sample has a larger mass fraction in the cold phase than the SF sample. Not only that but the cold phase extends to lower temperatures at large radii. The TNG Q halos, however, generally have less total gas than the TNG SF halos, so the larger cold gas fractions in the TNG Q sample do not necessarily imply that they have have a larger cold gas mass. Naively, one would assume that the presence of cold gas would be closely correlated with the star formation rate of the central galaxy, either due to star formation feedback driving winds that carry cold gas into the halo and/or because cold gas in the halo could be accreted and fuel star formation. Apparently the halo properties of the TNG quiescent galaxies are such that the cold phase at large radii is not accreted by the galaxy to fuel star formation. In the full TNG simulation quenching is associated with black hole feedback \citep{Weinberger+17, Nelson+19}, so it is likely that this mode of feedback is in part responsible for the cold phase properties. For example, the black hole feedback may be lifting cold material out of the ISM and launching it into the CGM \citep[e.g.,][]{Sanchez+19,Oppenheimer+20}, or stirring the CGM so violently that the cold gas cannot settle onto the central galaxy. Physically, the large cold fractions at large radii in the TNG Q and SF samples, and the \citetalias{Joung+12} simulation, relative to the idealized simulations may be associated with the larger velocity dispersions as shown in Figure~\ref{fig:all_profiles}.

The amount of cold gas at large radius and the physical mechanism by which it is produced differs amongst the idealized simulations. 
Broadly speaking, the condensation of cold gas in the idealized simulations is produced by one of two mechanisms. In the first mechanism cold gas is produced via highly mass loaded winds that eject significant amounts of gas into the inner CGM. This has a two-fold impact on the cold phase: it can directly launch cold material from the ISM into the CGM, and it can increase the halo density leading to more efficient cooling and multiphase condensation \citep[see][for a detailed description of this mechanism]{Li+20a}. The cold gas in the \citetalias{Fielding+17} high $\eta$, \citetalias{Su+20} Turb, and \citetalias{Li+20a} SFR3 simulations is primarily produced via this mechanism. Differences in the strength and implementation of the feedback in these three simulations results in the apparent differences in the amount and extent of the cold phase. 

The second mechanism that produces cold gas at large radii occurs when high specific energy outflows launch buoyant bubbles that lead to uplift of low entropy gas that efficiently cools and condenses at large radii \citep[see][for a detailed description of this mechanism]{LiBryan14,Voit+17}. The powerful high specific energy outflows in L20b SFR10 and the \citetalias{Su+20} Therm produce significant amounts of cold gas via this uplift mechanism. The \citetalias{Fielding+17} low $\eta$ simulation has a similarly high specific energy outflow but the total outflow is relatively weak due to the low star formation rate so the resulting cold phase is relatively minor. In general, the amount of cold CGM gas in the idealized simulations increases with the strength of the feedback regardless of which mechanism is dominant.

Stronger feedback leading to more pronounced and extended cold phases in the idealized simulations is not sufficient to explain the vastly higher mass fraction in, and larger radial extent of, the cold phase of the TNG simulations, nor the increasing prominence of the cold phase with radius in the \citetalias{Joung+12} simulation. Indeed, the idealized simulation with the most cold gas at large radii---the L20b SFR10 simulation---has a star formation rate many times that of the TNG SF sample. Additional physical processes must be at play that are not included in the idealized simulations. 

Possible processes include, but are not limited to, different feedback modes, filamentary accretion from the intergalactic medium, and the presence of satellite galaxies \citep{CAFG+15, CAFG+16, Hafen+17}. Feedback channels other than what are included in the idealized simulations can populate and maintain the cold phase (in particular the TNG two mode AGN feedback model). Filamentary accretion may provide cold gas directly into the CGM from outside. Satellites can both stir the CGM and introduce cold gas by stripping and wind ejection. The assembly of the halos over cosmic time, growing from small to large mass, may play a role in setting the phase structure, driving long lived turbulence, and setting up large scale velocity asymmetries in the outer halo that is not captured in the idealized simulations \citep[e.g.,][]{Vazza+11, Nelson+14}.

\subsection{Phase distribution}\label{sec:1D_phase}

\begin{figure*}
\centering
\includegraphics[width=\textwidth]{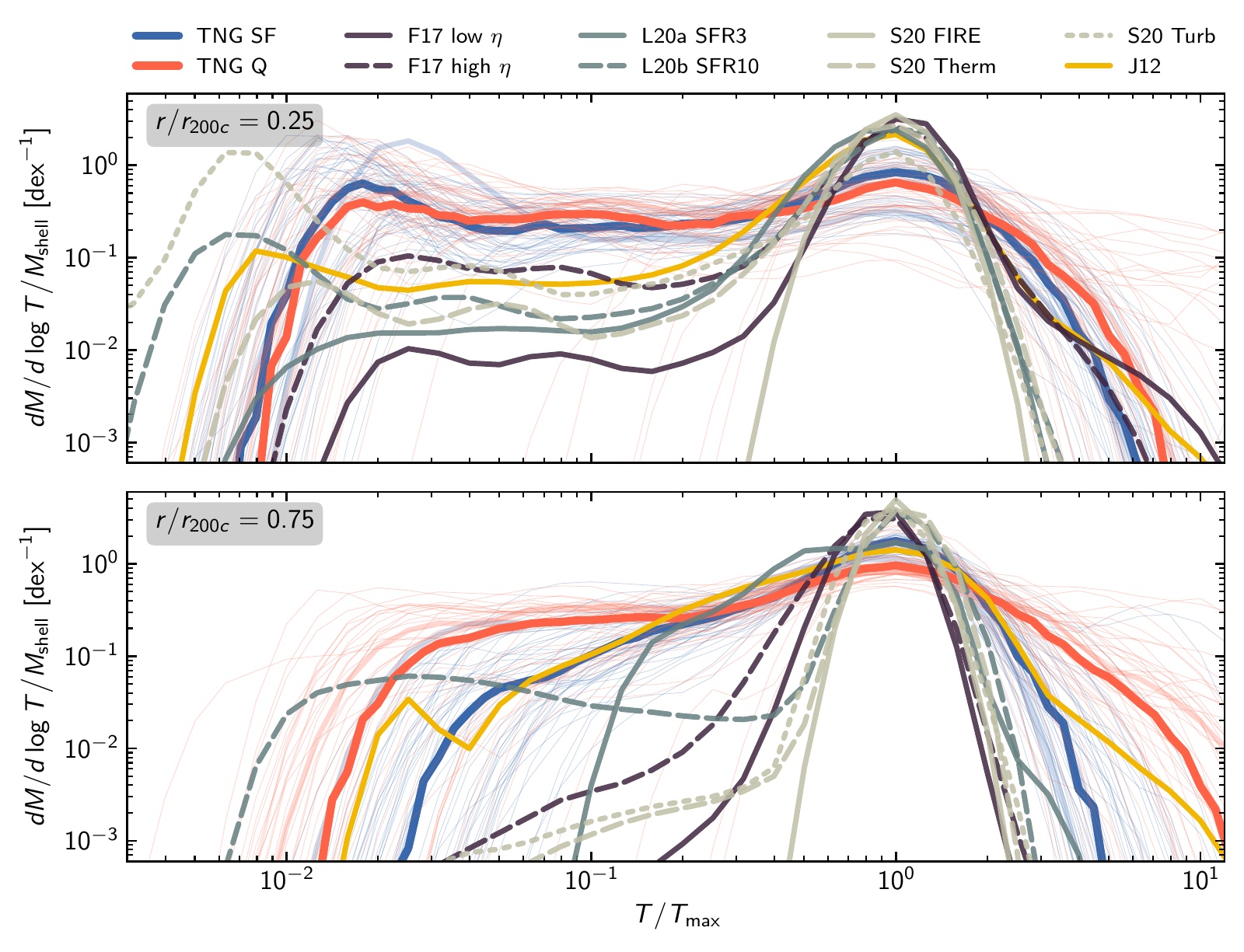}
\caption{The fractional temperature distribution (the amount of mass per logarithmically spaced temperature bin normalized by the total mass in the shell) in $0.1\,\rvir$ thick shells in the inner (centered on $0.25 \, \rvir$) and outer (centered on $0.75 \, \rvir$) CGM. The line styles are the same as in \autoref{fig:all_profiles}. The distributions are aligned to $T_{\rm max}$, the temperature where the distribution peaks, which is approximately the median temperature and is generally close to the virial temperature. This alignment highlights the relative shapes of the distributions rather than their normalizations. The idealized simulations have much more prominent virial components with significantly less low temperature material than the cosmological simulations---particularly in the outer halo. In the idealized simulations the amount of mass in the low $T$ phase is correlated with feedback strength. Whereas in the cosmological simulations the distributions are nearly flat and the quiescent galaxies' distributions are marginally broader on the high and low $T$ ends.}
\label{fig:dMdlogT}
\end{figure*}

The joint radial-temperature distributions examined in the previous section are useful to understand the overall distribution of gas phase within each simulation; however, to compare between simulations, it is useful, as we do in this section, to examine one dimensional phase distributions of various quantities within shells at a range of radii.

\subsubsection{Temperature phase profiles}
\autoref{fig:dMdlogT} shows the temperature distribution of the CGM mass fraction in the inner ($0.2-0.3~\rvir$) and outer ($0.7-0.8~\rvir$) halo. For reference, these shells are marked with arrows in the upper left panel of \autoref{fig:all_profiles} and the upper middle panel of \autoref{fig:all_phase}. To isolate the relative shapes rather than the differences in normalization (see \autoref{fig:all_profiles} for normalization differences) we have aligned the temperature distributions to the temperature value, $T_{\rm max}$, where each distribution reaches its maximum. In general $T_{\rm max}$ is close to $\Tvir$\footnote{In a few of the TNG halos, particularly in the inner CGM, $T_{\rm max}$ is in the cold phase, in which case we add the additional constraint that $T_{\rm max}$ must be in the hot phase ($T_{\rm max} > \Tvir /10$), which corrects the alignment.}. 

We can split the temperature distributions into three physically motivated components: the virial component, which is the hottest and encompasses gas near $T_{\rm max} \sim \Tvir$, the cold component, which is the coldest and corresponds to gas below $\lesssim T_{\rm max}/30$, and the intermediate component, which lies between the hot and cold phases and describes gas around $T_{\rm max}/30 - T_{\rm max}/2$.  In physical units the virial component traces gas at about $10^6$ K for halos in the mass range we are considering, the cold phase traces gas at about $10^4$ K, which is where gas in photoionization equilibrium with the meta-galactic UV background reaches thermal equilibrium, and the intermediate phase traces gas at about $10^5$ K, which should be the most transient/dynamic since this is the temperature around which the cooling rates are at their highest. 

At all radii, the virial components follow a roughly lognormal distribution that peaks at $T_{\rm max}$, albeit with significant deviations from lognormality. The width and the relative amplitude of this hot, virial component varies between different radii and simulations, however a few patterns emerge. Generally the virial phase is narrower and more prominent in the outer halo than in the inner halo. The virial phase (hot phase) is narrower and more prominent in the idealized simulations than in the cosmological simulations. There is a marked similarity between all of the idealized simulations' hot phase distributions at all radii. The \citetalias{Joung+12} simulation also agrees quite well with the idealized simulations in the inner halo, while at large radii it matches the TNG simulation distributions more closely. This points to the structure of the inner halo being predominantly set by feedback, whereas the structure of the outer halo is affected more by processes inherent to cosmological simulations.

The breadth of virial phase temperature distribution appears connected to the velocity dispersion. Regions with large velocity dispersions $\sigma_v$ (shown in the bottom right panel of \autoref{fig:all_profiles}) also have broad virial phase temperature distributions. At larger distances the narrowing of the idealized simulations' virial components when compared to the cosmological simulations is also reflected in the $\sigma_v$ profiles in \autoref{fig:all_profiles}. This highlights the same major discrepancy between the two simulation approaches that was seen in \autoref{fig:all_phase}, namely that there is a process, or processes (either physical or numerical) not included in the idealized simulations that leads to broader temperature distributions (and larger $\sigma_v$) at large radii in the cosmological simulations. This is also seen qualitatively in Figure \ref{fig:all_maps}.

The cold component further clarifies these differences. This cold component is centered at a temperature that ranges from about $T_{\rm max}/100$ to $T_{\rm max}/30$ when considering the inner or outer halo, respectively. This cold phase temperature is about $\sim 10^4$ K which is roughly where gas reaches thermal equilibrium with the UV background. This shifts relative to $\Tvir$ for different halos because of differences in halo mass, and increases with radius as the pressure drops and the equilibrium temperature rises. All simulations show the same radial trend: the cold, sub-virial component is more prominent in the inner halo.

As we focus on the outer shell in the bottom panel, the prominence of the cold component clearly differs between simulations.  It is virtually non-existent in the outer shell of all of the idealized simulations except in the L20b SFR10 simulation. On the other hand, in the cosmological simulations the cold phase persists out to large radii with only a small drop in prominence. As with the virial component, the fact that the \citetalias{Joung+12} AMR cosmological zoom-in simulation produces a cold phase that is similar to the idealized simulations in the inner CGM, but similar to the TNG simulations at large radii provides an important clue to the origin of cold gas in CGM simulations. The idealized simulations and the \citetalias{Joung+12} simulation have significantly higher resolution in the outer halo than the TNG simulations particularly in the volume filling component. Moreover the idealized simulations use both particle and grid based numerical methods. Therefore, the difference in outer halo cold component is likely not (only) a result of resolution or differences in numerical methods but most likely reflects a physical difference between the idealized and cosmological simulations. This strengthens the picture in which the inner CGM cold phase is regulated by feedback and the outer CGM cold phase is set by cosmological effects.

The intermediate temperature component, that traces the rapidly cooling phase near the peak of the cooling curve ($\sim 10^5$ K), forms a bridge between the hot, virial component and the cold component (when present). Its differential mass distribution is generally well described by a flat or a shallow power-law distribution. In all of the simulations the intermediate phase contains about as much mass as the cold phase. This is true both in the TNG simulations where the distinction between cold, intermediate, and hot phases is less well defined because the overall distributions are nearly flat, and in the idealized simulations where the hot component far dominates the mass budget. Physically, this temperature range is crucial to understand because it is where the cooling times are the shortest. Therefore, gas at these intermediate temperatures must be continually heated or this phase must be replenished at the same rate it cools out (unless it has very low density as in the case of an expanding hot outflow or the outermost regions of the halo). The fact that it contains as much mass per logarithmic temperature bin as the cold phase, which means there is not a pile up of gas at the coldest temperature, may provide important constraints on models for multiphase mixing and precipitation. 

Looking at the entire distribution, an additional pertinent detail is the overall breadth. In particular the median extent and shape of the TNG SF and Q samples agree quite well in the inner halo. The idealized simulations show some scatter in their breadth, with stronger feedback resulting in more low-temperature gas (as discussed with regards to \autoref{fig:all_phase}). When moving farther out in the halo, the gas in the TNG Q sample continues to have a broad hot phase distribution that extends out to significantly super-virial temperatures, whereas the TNG SF sample, as well as all of the rest of the simulations (idealized and cosmological), have essentially no gas beyond a few $T_{\rm max}$. Likewise, the cold phase of the TNG Q sample extends to lower temperatures than the cold phase of the TNG SF sample in the outer halo.  As we move to larger radii the amount of cold gas in the idealized simulations decreases until only the L20b SFR10 simulation shows significant cold gas.  


\begin{figure*}
\centering
\includegraphics[width=\textwidth]{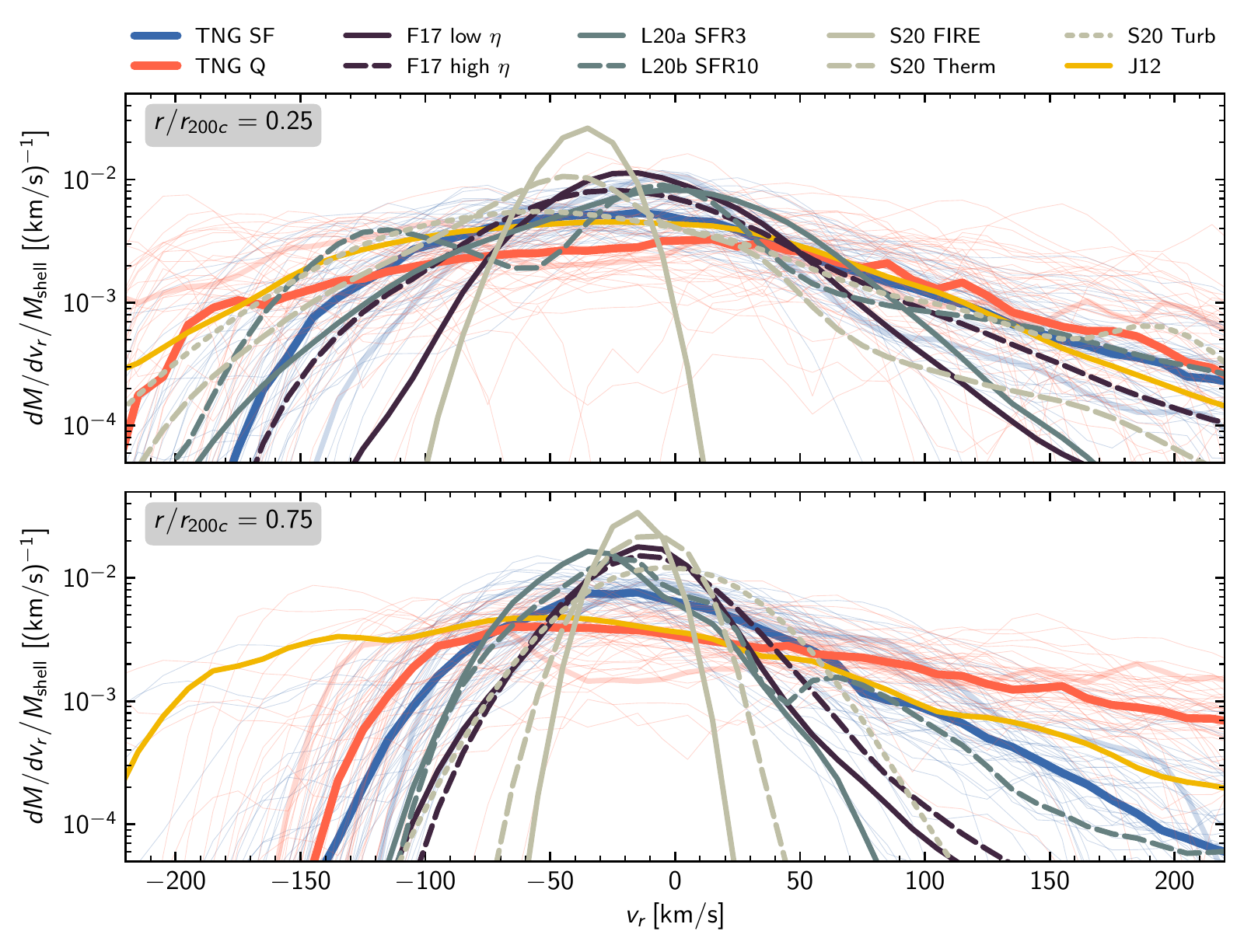}
\caption{The fractional radial velocity distribution (the amount of mass per radial velocity bin normalized by the total mass in the shell) in $0.1\,\rvir$ thick shells in the inner (centered on $0.25 \, \rvir$) and outer (centered on $0.75 \, \rvir$) halo, is shown in the top and bottom panels, respectively. Negative $v_r$ indicates an inflow toward the central galaxy. In the inner halo, the simulations all peak at some small negative value ($\sim -20$ km/s), and have similar breadths with the exception of a few of the weaker feedback idealized simulations. In the outer halo, the cosmological simulations have significantly broader distributions on both the low and higher velocity end than any of the idealized simulations. The L20b SFR10 has a high velocity tail comparable to the cosmological simulations, and the \citetalias{Joung+12} has a large inflowing component consistent with the TNG halos with the largest inflows.}
\label{fig:dMdvr}
\end{figure*}

\subsubsection{Velocity phase profiles}
Unlike the temperature distributions the inner and outer halo radial velocity distributions, shown in the top and bottom panels of \autoref{fig:dMdvr}, respectively, do not exhibit distinct components. Instead the distributions are smooth and unimodal. In general the distributions have a negative median around $\sim -20$ km/s, which corresponds to a moderate inflow. The distributions are asymmetric about this median. The inflowing wing falls off steeply beyond $\sim -$150 km/s $\approx -\vvir$, whereas the outflowing wing has a broad tail that extends to high velocities $\gtrsim 200$ km/s in some cases. 

In the inner halo there is a broad similarity across all of the simulations with the noted exception of the \citetalias{Su+20} FIRE simulation, which has a significantly narrower and more symmetric radial velocity distribution that reflects its lack of appreciable outflows. The cosmological simulations and the strongest feedback idealized simulations have comparable outflowing radial velocities. This reflects the importance of feedback in setting the structure of the inner CGM and highlights the utility of idealized simulations for studying the interplay of galactic winds and the CGM. In the idealized simulations the lower momentum wind models lead to narrower velocity distributions. The TNG Q sample and the \citetalias{Joung+12} halo have more rapidly inflowing material than the TNG SF sample. The large inflows in the quiescent population is at first surprising, but when comparing to the \citetalias{Su+20} Turb simulation we can see that the full velocity distributions are comparable. It is, therefore, possible (although by no means certain) that the large inflow velocities are not representative of a coherent flow, but instead represent a component of a nearly randomly distributed velocity field. This is likely true to some degree for all of the distributions, and is corroborated by the relatively small inflowing median velocity and the large velocity dispersion as shown in the bottom left and right panels of \autoref{fig:all_profiles}, respectively \citep[see][for more discussion on this point]{Lochhaas+20}. However, we note that the velocity flows highlighted in \autoref{fig:all_maps} seem to indicate that the cosmological simulations have more large-scale coherent flows that are asymmetrically distributed rather than exhibiting smaller scale motions found evenly around the CGM. 

Focusing on the inflowing side of the velocity distribution, when going from the inner to the outer halo the extent of the inflow velocity distribution decreases. This is consistent with a picture in which the highest velocity inflowing material is in free-fall from large radii, so the left edge of the distribution roughly traces the free-fall velocity at that radius \citep[e.g.,][]{ForbesLin19,Mandelker+19}. The \citetalias{Joung+12} simulation, along with some of individual TNG halos, represent extreme cases where there are nearly super-virial inflow velocities at large halo-centric radii (we also remind readers that the results for these simulations represent a single snapshot and are not time-averaged). In the outer halo the cosmological simulations have appreciably more inflowing gas than the idealized simulations. This broad distribution is likely a reflection of the large scale velocity asymmetries that are apparent in \autoref{fig:all_maps} and that arise due to the cosmological assembly of these halos.

The outflowing side of the inner halo velocity distributions of the idealized simulations with the strongest winds and the largest momentum content have high velocity tails quite similar to that of the cosmological simulations. Idealized simulations with weaker winds have less material moving at the highest outward velocities ($\gtrsim 100$ km/s). Farther out in the halo the high velocity outflowing material falls off in all but the TNG Q sample, which may be a reflection of the powerful AGN feedback emanating from the TNG Q galaxies. In the idealized simulations this high velocity material disappears more rapidly than in the cosmological simulations, and the full radial velocity distributions become roughly symmetric about the median in the outer halo. Only L20b SFR10, which has the most powerful winds of the idealized simulations, shows an extended outflowing velocity tail in the outer halo. High velocity outflowing material persists out to large radii in the cosmological simulations, especially the TNG Q sample, in which the mass fraction above $200$ km/s actually increases in the outer halo relative to the inner halo. Given that the idealized simulations presented here span a broad range of feedback properties it is unlikely that feedback alone is responsible for the differences between the idealized simulations and cosmological simulations in the outer halo. In the inner halo, however, feedback appears to be the predominant mechanism controlling the velocity distribution shape.


\begin{figure*}
\centering
\includegraphics[width=\textwidth]{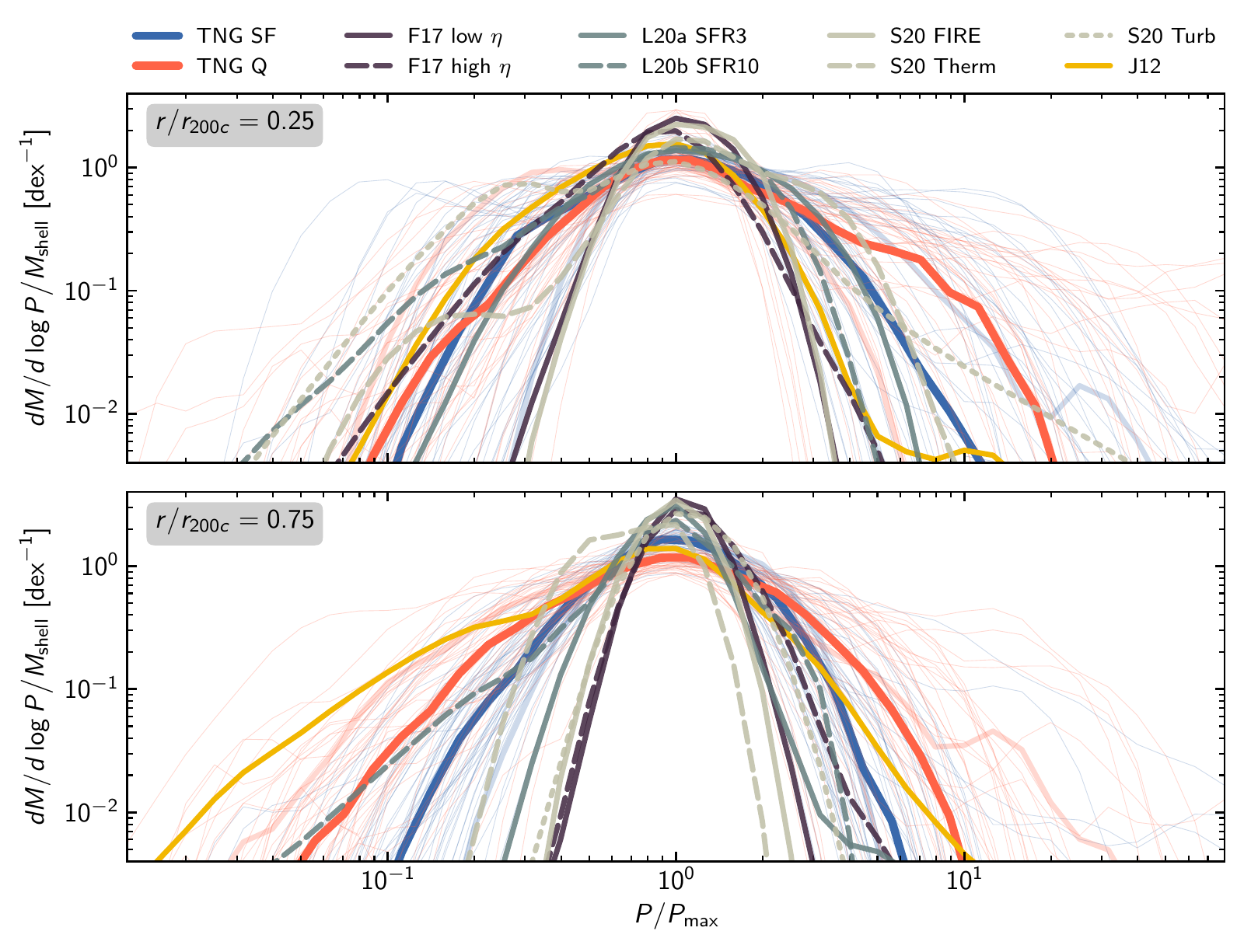}
\caption{The top and bottom panels show the fractional pressure distribution (the amount of mass per logarithmically spaced pressure bin normalized by the total mass) in $0.1\,\rvir$ thick shells in the inner (centered on $0.25 \, \rvir$) and outer (centered on $0.75 \, \rvir$) CGM, respectively. The distributions are aligned to $P_{\rm max}$, the pressure where there is the most mass, which is approximately the median pressure. This normalization highlights the relative shapes of the distributions rather than their alignment. The idealized simulations have systematically narrower pressure distributions than the cosmological simulations---particularly farther out in the halo. The quiescent TNG sample has a marginally broader distribution than the star forming TNG sample. }
\label{fig:dMdlogP}
\end{figure*}

\begin{figure*}
\centering
\includegraphics[width=\textwidth]{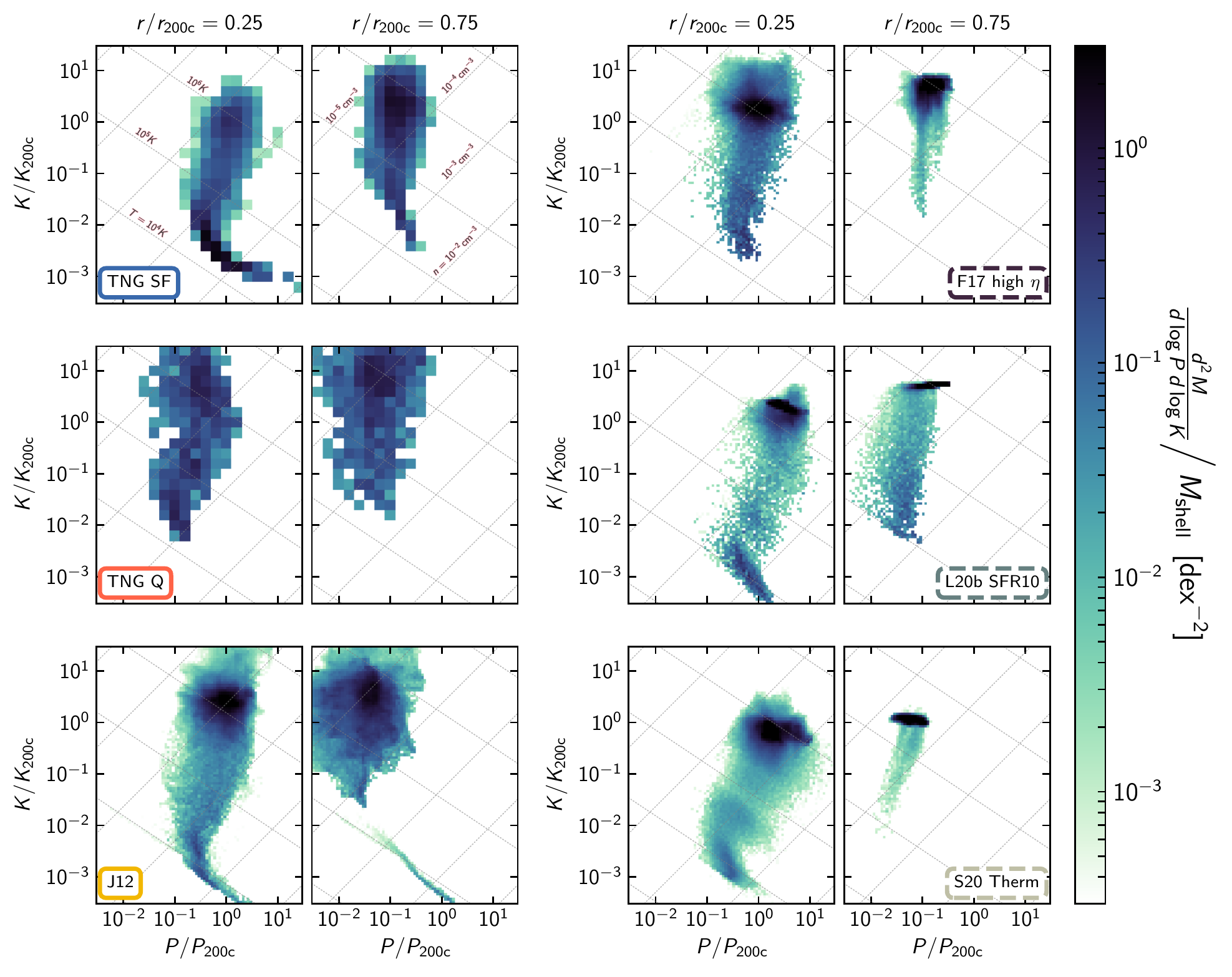}
\caption{The pressure-entropy phase diagram in $0.1\,\rvir$ thick shells in the inner (centered on $0.25 \, \rvir$) and outer (centered on $0.75 \, \rvir$) CGM. The left column shows the distributions in single representative TNG SF and Q halos (same halos as in \autoref{fig:all_maps} and \autoref{fig:all_phase}) and the \citetalias{Joung+12} halo. The right column shows distributions in the three idealized simulations that have the most cold gas at large radii: \citetalias{Fielding+17} high $\eta$, L20b SFR10, and \citetalias{Su+20} Therm. The TNG halos are binned using coarser bins because of their lower resolution. The thin dotted lines show lines of constant temperature ($K\propto P^{-2/3}$) and constant density ($K\propto P$). All simulations exhibit a clear hot, virial component at entropies around $\sim 3 \Kvir$ and tails that extend to low entropy. In the inner halo the distributions exhibit a shift to lower pressure at intermediate entropies ($\sim 10^{-2} \Kvir$). At low entropy ($\sim 10^{-3} \Kvir$), the pressure increases following a constant $\sim 10^4$ K temperature track up to, or above, the level of the hot phase. In the outer halo this intermediate-entropy pressure decrement is generally less pronounced.}
\label{fig:pressure_entropy}
\end{figure*}

\subsubsection{Pressure phase distribution}
\autoref{fig:dMdlogP} shows the pressure distributions of the mass fraction in the inner and outer halo. 
As with the radial velocity, the distributions in both shells are roughly lognormal, although they have different widths and asymmetric tails to high and low $P$.
As with the temperature distributions, the width of the pressure distributions appears correlated with the magnitude of the velocity dispersion. Wider pressure distributions occur in regions with large velocity dispersions, which may be a result of large scale asymmetries or small scale turbulent fluctuations. 

Deviations from lognormality on the high pressure end are likely due to over pressurized bubbles and shocks that arise during strong feedback events. On the low pressure end, deviations from lognormality may arise from rapidly expanding bubbles temporarily out of equilibrium, or in rapidly cooling regions in which cooling proceeds isochorically. Runaway rapid cooling is notoriously hard to resolve numerically and under-resolved multiphase condensation often proceeds isochorically when in the limit of infinite resolution it would proceed isobarically and thereby cause no deviation to the pressure distribution \citep[e.g.,][]{Fielding+20}. When the multiphase condensation does not have sufficient resolution, the condensing cold clouds contract down to the resolution limit prior to saturating and cannot break up into smaller clumps that would remain in sonic contact \citep[see for example,][]{McCourt+18, Nelson+20}. Isochoric cooling can, of course, also occur in fully resolved scenarios, but because of its out-of-equilibrium nature it is expected be transient and less common.  

The pressure distributions of the idealized simulations are broader in the inner halo than in the outer halo, which is similar to the decrease in velocity dispersion at large radii as shown in \autoref{fig:all_profiles} and \autoref{fig:dMdvr}. In the inner halo the most turbulent idealized simulations with the strongest feedback have pressure distributions of similar width as the cosmological simulations, while the idealized simulations with weaker feedback have appreciably narrower distributions. In the outer halo, the idealized simulation with the largest cold gas fraction (L20b SFR10) has a pressure distribution roughly as broad as the TNG SF sample, which is itself narrower than in the TNG Q and \citetalias{Joung+12} halos. The broader pressure distributions, which occur in regions with large velocity dispersions, may be linked to the origin of the cold and intermediate temperature phases seen in \autoref{fig:all_phase} and \autoref{fig:dMdlogT} since it is the tails of the distribution that are most susceptible to cooling out. 

\subsection{Joint Pressure-Entropy distribution}\label{sec:2D_phase}

For a final point of comparison we show in \autoref{fig:pressure_entropy} the joint pressure-entropy distribution in the inner ($0.2 \leq r/\rvir \leq 0.3$) and outer ($0.7 \leq r/\rvir \leq 0.8$) CGM of two representative TNG halos from the SF and Q samples (same halos as shown in \autoref{fig:all_maps} and \autoref{fig:all_phase}), the \citetalias{Joung+12} simulation, and the three idealized simulations that have the largest cold phases: \citetalias{Fielding+17} high $\eta$, L20b SFR10, and \citetalias{Su+20} Therm. The four idealized simulations not shown are similar to the three shown idealized simulations but are more dominated by a compact virial phase distribution and have smaller tails down to low entropies. The thin dotted gray lines trace constant temperature ($T=10^4$, $10^5$, and $10^6$ K normalized appropriately for a $10^{12} M_\odot$ halo; $K \propto P^{-2/3}$) and constant number density ($n=10^{-5}$, $10^{-4}$, $10^{-3}$, and $10^{-2}$ cm${}^{-3}$; $K\propto P$). 

The pressure-entropy space (as opposed to density-temperature, for example) is the most convenient to work in because the dominant physical processes in the CGM are isobaric and isentropic (rather than isothermal and isochoric). Cooling causes gas to move down the entropy axis, whereas shocks and heating move gas up the entropy axis. Adiabatic processes instead move gas along the pressure axis while keeping the entropy the same. Important adiabatic processes in the CGM include expansion and compression. Expansion leads to a loss of pressure and often occurs within wind bubbles as they move out into the halo. Compression leads to a pressure increase and often occurs as ambient material is swept up during violent feedback events. Turbulence naturally produces both expansion and compression and, therefore, leads to an overall broadening of the pressure distribution.\footnote{Shocks and the dissipation of turbulent motions do, however, increase the entropy, especially when the turbulence is supersonic.} Thus, trends in the pressure-entropy distribution are useful in elucidating the dominant physical and numerical processes shaping the CGM. The informative details revealed in \autoref{fig:pressure_entropy} are, however, subtle and are not captured in less granular analysis, such as one-dimensional distributions (\autoref{fig:dMdlogT}, \autoref{fig:dMdvr}, and \autoref{fig:dMdlogP}) and median radial profiles (\autoref{fig:all_profiles}).  

Fully establishing what determines the shape of the pressure-entropy distributions is beyond the scope of this comparison project. As such, we explicitly restrict our speculation about the possible reasons underlying the distributions' shapes to general connections to relevant phenomena. Instead, we focus primarily on an empirical description of the distributions in the inner and outer halo of the idealized and cosmological simulations. 

In addition to further supporting the trends regarding the amount of cold gas and the width of the pressure distribution discussed above, \autoref{fig:pressure_entropy} highlights intriguing commonalities and differences in the details of the distribution of CGM gas in these simulations. We first highlight the similarities across the simulations.  The simulations all exhibit a high entropy ($K\gtrsim \Kvir$), virialized component in the inner and outer halo. This high entropy node has a tail extending to low entropies ($K\lesssim 10^{-3} \Kvir$). At intermediate entropy values ($K \sim 10^{-2} \Kvir$), the pressure shifts below the value of the high entropy, virial component. This decrement is more pronounced in the inner halo than in the outer halo in all simulations. At low entropy the pressure increases back toward the pressure of the high entropy, virial component following a ${\sim} 10^4$ K isothermal contour that corresponds to thermal equilibrium with the meta-galactic UV background.

Despite the similarities in the general shape of the $P$-$K$ phase diagram, intriguing variations point to different physical processes in the simulations.  The magnitude of the inner halo intermediate-entropy pressure decrement varies between the simulations. The L20b SFR10 and \citetalias{Su+20} Therm halos have a decrement of more than an order of magnitude, the \citetalias{Fielding+17} high $\eta$ and \citetalias{Joung+12} have a decrement of ${\sim} 3$, and the TNG halos' decrements fall between these limits.  The outer halo of the cosmological simulations show nearly constant pressure more like \citetalias{Fielding+17}, however, while L20b SFR10 and \citetalias{Su+20} Therm halos still show a small decrement. In the outer halo of the \citetalias{Joung+12} simulation there is a conspicuous ${\sim} 10^4$ K tail that extends to much higher pressure than the main virial component. This tail is likely a satellite galaxy or filament, which is over pressurized due to (self-)gravitational confinement. 

This dual pressure-entropy view of the simulations motivates a deeper look into the cause of the intermediate-entropy pressure decrement. Two plausible, although by no means exclusive, explanations for this pressure decrement are (i) adiabatic expansion in a wind and (ii) numerically unresolved cooling. As discussed above, high specific energy outflows efficiently inflate wind bubbles that adiabatically expand as they the sweep up material. This uplift naturally promotes multiphase condensation, and in the process the wind material will lose pressure as it expands. This may explain why the idealized simulations with high specific energy outflow winds, L20b SFR10 and \citetalias{Su+20} Therm, have larger pressure decrements than the \citetalias{Fielding+17} high $\eta$ simulation, which has lower specific energy outflows (see discussion in \autoref{sec:temperature_distribution}). 

The intermediate-entropy pressure decrement may also be due to unresolved cooling. \cite{Fielding+20} demonstrated that cooling in under-resolved multiphase gas leads to pressure decrements at intermediate entropy where the cooling time is the shortest that are similar to what is shown in \autoref{fig:pressure_entropy} \citep[see also][for discussion of similar pressure decrements in under-resolved thermal instability simulations of the ISM]{Piontek+04,JGKim+13}. This under-resolved multiphase cooling can lead to significant errors in the total cooling rate. All of the simulations we analyze here are under-resolving the CGM cold phase to some degree, so this effect is likely present at some level. 

We encourage future studies to investigate the underlying cause (physical or numerical) of the intermediate-entropy pressure decrements found here. An important clue on the origin of this feature in the pressure-entropy distributions may be found in the relative diminution of the decrement at large radii where the impact of outflows and cooling is diminished, and where the resolution is generally lower. These decrements are not only tied to important physical processes shaping the CGM, but also fall in a range commonly probed by quasar absorption line observations. 

\section{Discussion}\label{sec:discussion}

This paper has presented direct comparisons of the CGM properties of several different simulations.  Here we discuss the utility of analyzing idealized and cosmological simulations in combination (Section \ref{sec:ideal_vs_cosmo}), and then highlight one of the main conclusions this comparison has allowed (Section \ref{sec:origin}).  Importantly, we connect the simulations to observations, and discuss the caveats to our work in the final two subsections.

\subsection{Idealized versus Cosmological}\label{sec:ideal_vs_cosmo}
There are clear advantages and limitations to both the idealized and cosmological simulations.

The idealized simulations we have analyzed simulate a volume ${\sim}10^6$ times smaller than the cosmological simulations. The idealized simulations are therefore able to achieve higher spatial/mass resolution and can output at high frequency with shorter intervals, all at a fraction of the computational cost. Because of the controlled nature of these numerical experiments, multiple feedback prescriptions can be applied to otherwise identical halos. Moreover, the feedback models employed were chosen based on considerations of the underlying physics launching the outflows. The simplicity of the idealized simulations' designs and the limited set of included physical processes allows for identifying and understanding the dominant mechanisms responsible for emerging phenomena. 

The benefits of idealized simulations, however, come at the cost of realism and may oversimplify to the point of missing essential ingredients. In particular, the lack of the cosmological context, the simple spherical initial conditions, and the small sample sizes limit the overall utility of idealized simulations.

Cosmological and cosmological zoom-in simulations consider a much larger volume that evolves from cosmological initial condition. This naturally results in more realistic inflows into the halos, including the accretion of satellite galaxies. The TNG simulation has larger samples of systems at specific galaxy/halo masses, which enables the variance among halos to be studied. 

However, cosmological simulations generally require vastly higher computational expense than the idealized simulations. Moreover, the feedback models employed are generally not physically motivated and are instead parameterized in an ad hoc fashion for galaxies to match certain observables. A result of the large volumes simulated is that the resolution is poor in the CGM, which can lead to inaccurate predictions for the phase structure and dynamics. Lastly, because of the high degree of physical realism that is sought by including many physical processes simultaneously, it is usually not straightforward to trace back the physical mechanism for certain phenomena.

The two cosmological simulations we have analyzed here represent only a small fraction of the diverse models used in studying galaxy formation in cosmological simulations. In particular feedback models used in cosmological simulations span a broad range of incarnations that have disparate impacts on galaxies and the gas that surrounds them. \cite{Davies+20}, for example, demonstrated that the halos in the EAGLE and TNG simulations have very different median CGM mass fractions in the halo mass range we have considered, which is likely a result of their very different AGN feedback models. Nevertheless, the EAGLE simulations support our general finding that cosmological simulations have significant velocity dispersions ($\sigma_v/\vvir \gtrsim 0.5$ at $\rvir$) at large radii \citep{Oppenheimer18} and broad temperature distributions \citep{Oppenheimer+18} are also found in the EAGLE halos.

\subsection{Origin of Cold Gas in the Inner and Outer Halo}\label{sec:origin}

Our comparative analysis of the cosmological and idealized simulations supports a picture in which the cold gas of low-redshift Milky Way mass galaxies in the inner ($\lesssim 0.5 \rvir$) and outer CGM ($\gtrsim 0.5 \rvir$) has different origins. In the inner halo, galactic feedback is responsible for cold gas production via processes such as uplift and direct injection. In the outer halo, cold gas is a result of inherently cosmological processes as opposed to feedback from the central galaxy. 

By comparing the \citetalias{Joung+12} adaptive-mesh-refinement cosmological zoom-in simulation to the other simulations, the differing mechanisms for generating cold gas in the inner and outer halo become clear. The \citetalias{Joung+12} halo has a similar inner CGM temperature distribution to that of the idealized simulations, but this differs significantly from the TNG SF and Q halos. Moving to the outer CGM, the temperature distributions of the idealized simulations diverge even more from the TNG halos, whereas the \citetalias{Joung+12} temperature distribution becomes consistent with the TNG halos. This points to a process omitted from the idealized simulations that promotes the production and/or maintenance of cold gas in the outer halo, as seen in the cosmological simulations. This cold gas excess is also associated with broader pressure and velocity distributions and large velocity dispersions in the outer CGM of the cosmological simulations. The most obvious ingredients that are present only in the cosmological simulations include: the existence of inflowing low-entropy filaments, the presence of satellites, and the hierarchical assembly of halos. The fact that the \citetalias{Joung+12} cosmological simulation has an inner CGM cold phase (and other properties) similar to the idealized simulations indicates that the differences with the inner TNG halos is not a result of inherently cosmological effects. These differences are instead a result of the powerful feedback models, including AGN feedback, employed in the TNG simulation (the \citetalias{Joung+12} feedback is relatively weak).

This finding is supported by the recent in-depth analysis of the CGM of massive galaxies in the TNG50 simulation (the higher resolution, smaller volume counterpart to the TNG100 simulation) by \citet{Nelson+20}. They found that cold gas in the CGM---as traced by H\textsc{i} and Mg\textsc{ii}---formed via thermal instabilities that were seeded by large density perturbations. Feedback is the dominant perturber in the inner halo of both cosmological and idealized simulations. The idealized simulations analyzed here have no mechanism to generate large perturbations in the outer halo other than feedback. The cosmological simulations, on the other hand, have many channels besides feedback to seed the necessary perturbations as a result of their hierarchical growth over cosmic time. 

\subsection{Ingredients for an Idealized CGM}
The missing cold gas formation channel in the outer halos of idealized simulations presents an opportunity for future experiments to definitively identify what physical processes are responsible for the extensive cold phase in cosmological simulations. 
The dominant formation channel may be uncovered by incrementally adding additional processes to idealized simulations. 
The enhanced cold phase at $r/\rvir \lesssim 0.2$ of the \citetalias{Su+20} simulations relative to the \citetalias{Fielding+17} and \citetalias{Li+20a},b simulations has, for example, highlighted the impact of including the rotation of the halo gas \citep[see][for a detailed study of the impact of rotation on the CGM]{DeFelippis+20}.

The challenge for the next generation of idealized CGM simulations is to include additional processes in a controlled fashion so as to clearly identify the underlying cause of the growth of the cold phase (or any other changes that may manifest as processes are added). How to sensibly include satellites, substructure, different AGN feedback models, cosmological accretion, or the evolution of the dark matter halo is non-trivial, but the utility in providing a complement to cosmological simulations is key to unlocking an intuitive physical model for the nature of halo gas. 

\subsection{Connecting to Observations}

Although this work has been focused on comparing the physical properties between simulations, here we briefly comment on possible comparisons with observations.  

The most reliable statements we can make pertain to the hot, volume-filling phase of the CGM.  This is because the hot gas is the most well-resolved across all simulations, and also is less affected by differences in the implemented cooling rates.

We can roughly use the temperature of our hot halos to estimate where oxygen ions---a common observational probe---might be observed in our CGM simulations.  Specifically, when collisionally ionized, O VI requires gas at 10$^{5.5}$ K, O VII exists between 10$^{5.5}$-10$^{6.5}$ K, and O VIII is found at the highest temperatures of 10$^{6.2}$-10$^{6.7}$ K.  All of the halos follow a similar temperature profile, with O VIII-temperature gas being found in the most central regions and more O VI able to be produced closer to the halo outskirts where the temperatures are lower.  The temperature profiles are, however, not identical, and differences in the normalization and shape of the temperature profiles will lead to differing amounts of gas in the $\sim 10^{5.5}$K range that is traced by O\textsc{vi}. A more rigorous comparison of column densities is required to determine the different observational predictions across the simulations, which we leave to an upcoming work. 

Predictions for observations sensitive to the cold, clumpy phase of the CGM from the set of simulations we have analyzed here are highly uncertain given the disparity in cold phase properties in the cosmological and idealized simulations. Nevertheless, what our results---particularly \autoref{fig:all_phase} and \autoref{fig:dMdlogT}---indicate is that the ratio of ions tracing the cold phase (e.g., H\textsc{i}, Mg\textsc{ii}, Si\textsc{ii}, C\textsc{ii},...) to ions tracing the intermediate (e.g., N\textsc{iii}, C\textsc{iv}, Si\textsc{iv}) and hot phases (e.g., N\textsc{v}, O\textsc{vi}, Ne\textsc{viii}) will differ dramatically between the idealized and cosmological simulations. In all of the idealized simulations at all radii the fraction of mass in the hot phase far dominates the intermediate and cold phases, whereas, in the cosmological simulations, there is almost equal mass in all phases. Observed ion ratios can, therefore, be used to distinguish between the different temperature distributions and provide stringent constraints on the underlying physics regulating these important phases.

In addition to this rough comparison of expected ion column density predictions, it is clear that the idealized and cosmological simulations would lead to markedly different kinematic signatures. The cosmological simulations have significantly larger velocity dispersions at large radii (see \autoref{fig:dMdvr}), which would lead to significantly broader line profiles than in the idealized simulations.

\subsection{Caveats}

\subsubsection{Comparing Different Halos}

None of the simulations included in our analysis were run with the express purpose of comparing to other simulations in the rigorous manner that we have attempted here. Therefore there are differences that may not be entirely negligible. 

The dark matter halo mass of all of our simulations differ by up to a factor of 2. In most of our analysis we have looked at virial normalized quantities, which should in theory scale out any halo mass differences. Cooling and feedback, however, introduce a scale dependence that is not captured by this normalization process, so some apparent differences in the halo gas properties may be due to this halo mass mismatch. Moreover, the shape of the dark matter profiles differ. In particular, the circular velocity profile is more centrally peaked in the \citetalias{Joung+12}, \citetalias{Su+20}, and \citetalias{Li+20a},b halos than in \citetalias{Fielding+17} halos, which are nearly flat. The TNG halos are roughly between these two limits. As a result, it is likely that the dark matter profile differences are responsible for the differences in the shape of the temperature profiles of the virial component (see \autoref{fig:all_profiles}). 

In addition to differences in the dark matter halos and feedback models, the initial conditions and included processes also differ amongst the idealized simulations. The \citetalias{Su+20} simulations start with a nearly baryonically complete hot gaseous halo out to the edge of the simulation domain. The \citetalias{Li+20a},b simulations start with a low density ambient halo and rely on winds from the central galaxy to populate the halo. The \citetalias{Fielding+17} simulations start with a hydrostatic hot halo out to $\sim 0.7 \rvir$ that contains less than the cosmic baryon budget. The \citetalias{Fielding+17} simulations, however, also include spherical accretion from large radii to mimic the growth of the halo over time, which increases the CGM mass up to baryon completeness after $\gtrsim 6$ Gyr, which is when the analysis begins. The \citetalias{Su+20} simulations include rotation of the halo gas while the \citetalias{Fielding+17} and \citetalias{Li+20a},b do not. The idealized simulations all use different treatments of radiative cooling and metallicity. 

In this work, we find that the most interesting difference between the various idealized simulations examined is the feedback model that each work adopts. \citetalias{Li+20a},b assumed constant star formation rates and launched winds with mass and energy loadings calibrated to small scale resolved ISM simulations. \citetalias{Fielding+17} on the other hand allowed the star formation rate to self consistently vary according to the galactic accretion rate and used mass and energy loadings broadly commensurate with what is found in the FIRE simulations \citep{Muratov+15,Angles-Alcazar+17}. The \citetalias{Su+20} simulations used the ISM models from the FIRE-2 simulations to self consistently form stars and drive winds, and added additional energy sources to represent the possible impact of AGN feedback. These differences undoubtedly impact the quantitative details of our findings in hard-to-isolate ways, but taken as a whole they strengthen our primary qualitative finding that none of the processes included in the idealized simulations can reproduce the broad phase structure and large velocity dispersions found at large radii in the cosmologically simulated halos that we study.

\subsubsection{Resolving the Phases of the CGM}

The most robust conclusions that can be drawn from our analysis pertain to the hot phase of the CGM because it is well resolved in all of the simulations we have looked at. Conclusions about the cold phase, as we have stressed throughout, are less sure because it is unlikely that the cold phase is well-resolved in any of the simulations presented here. The consequences of under-resolving the cold phase is unclear and a topic of ongoing research \citep[e.g.][]{Fielding+20}.  Recently, cosmological simulations with novel methods for improving the resolution in the CGM have demonstrated the dramatic sensitivity of observational predictions, particularly for tracers of cold gas, on the CGM resolution \citep{vandeVoort+19,Hummels+19,Peeples+19,Suresh+19,Mandelker+19b}. Likewise, \citet{Nelson+20} demonstrated, using a detailed analysis of the TNG50 simulation, that the CGM cold phase, primarily comprised of distinct clouds that formed via thermal instabilities, are highly sensitive to numerical resolution. Unresolved cooling may lead to the lack of pressure equilibrium as seen in \autoref{fig:pressure_entropy}. It may have far reaching consequences on the growth and evolution of the galaxies. 

\subsubsection{Physics not included}

None of the simulations discussed here include thermal conduction or cosmic rays. Thermal conduction is known to play a major role in setting the phase structure in the ISM \citep[e.g.,][]{Cowie1977} and could be equally important for the CGM. It is unlikely the conduction will affect the overall CGM structure \citep{Su+17,Hopkins+20} because the conduction timescales are quite long in the virialized phase of halos with this mass. However conduction could be especially important when it comes to determining the details of the cold phase mass distribution. 

Recently, the impact of cosmic rays on the structure and evolution of the CGM has received much interest \citep[e.g.,][]{Salem+16, Ji+20, Buck+20}. While these preliminary investigations are still quite rough given the large uncertainty in cosmic ray transport \citep{Hopkins+20b}, they point to a picture in which the cosmic ray energy density may play a major role in supporting cool material at large radii. When cosmic ray pressure is large relative to thermal pressure it can cause gas to cool isochorically, which causes dramatic changes to the phase structure of the CGM. As more sophisticated treatments of cosmic ray transport on galactic and intergalactic scales are explored the resulting changes to the CGM properties will have important physical and observational implications.

\section{Conclusions}\label{sec:conclusion}

We have presented a comparative analysis of the CGM properties in seven CGM-focused idealized simulations \citep{Fielding+17,Su+20,Li+20a}, a cosmological zoom-in simulation \citep{Joung+12}, and in two sets of 50 star forming and quiescent galaxies from the TNG100 cosmological simulation \citep{TNG-release}. By analyzing this diverse set of simulations in a uniform manner we have been able to isolate commonalities and differences in the CGM. 
Our results show how the median CGM properties (temperature, density, pressure, entropy, radial velocity, and velocity dispersion) scale with radius (see \autoref{fig:all_profiles}). Additionally, we investigated how the shape of temperature, radial velocity, and pressure distributions vary about the median values in the inner and outer CGM (see \autoref{fig:all_phase}-\ref{fig:dMdlogP}). Lastly, we use the joint pressure-entropy phase distribution in the inner and outer halo in a subset of our simulation sample to highlight instructive, albeit subtle, CGM properties that are closely tied to the underlying physical processes and numerical methods (\autoref{fig:pressure_entropy}).

The median properties highlighted that, for the most part, the temperature is within a factor of a few of $\Tvir$. Differences in the shape of the median temperature profiles are due in part to differing multiphase distributions and to differences in the underlying dark matter distributions. The median density distributions are similar throughout most of the halo with the notable exception of the TNG SF sample which has a factor of 2-3 more baryons in the halo than any of the other halos. The median velocity dispersion is significantly higher at all radii (particularly large radii) in the cosmological simulations than in the idealized simulations. A visual comparison presented in \autoref{fig:all_maps} indicates that the large velocity dispersion is due primarily to halo-scale velocity asymmetries.

The difference between the idealized and cosmological simulations is more pronounced when looking at the mass distributions as opposed to the median values. Although all of the simulations have unimodal pressure and radial velocity distributions, the cosmological simulations have systematically broader distributions in the outer halo. The temperature distributions showed the most noticeable differences between the idealized and cosmological simulations. The idealized simulations all exhibit a prominent hot, virial phase that contains at least an order of magnitude more mass than the intermediate and cold phases, and the cold phase mostly disappears in the outer halo. In the idealized simulations the prominence of the cold phase depends sensitively on the feedback model. By contrast, The TNG SF and Q halos have nearly as much cold and intermediate temperature material as in the hot phase, and the cold phase remains significant throughout the halo. The inner CGM temperature distribution of the \citetalias{Joung+12} halo is similar to that of the idealized simulations, but in the outer CGM it is similar to that of the TNG halos.

The clear differences in CGM properties between the idealized, cosmological, and cosmological zoom-in simulations highlight that different physical processes shape the inner and outer CGM. Feedback from the central galaxy is primarily responsible for determining the inner CGM phase structure. Feedback promotes multiphase condensation and sustains cold gas in the inner halo via direct injection and/or uplift. Inherently cosmological effects, which are not included in the idealized simulations, are responsible for the broad outer CGM temperature, velocity, and pressure distributions that are present in all of the cosmological simulations but are absent in the idealized simulations. The most likely drivers of this difference are the presence of satellite galaxies and nonspherical cosmological accretion that evolves in time. This poses a challenge for future idealized simulations to develop methods to capture the crucial outer CGM cosmological flow interactions in order to expand the usability of the inherently efficient idealized CGM simulations out to larger distances.

The huge diversity of CGM attributes in our heterogeneous simulation sample---particularly in the highly feedback dependent inner CGM---underscores the uncertainty in the true state of the multiphase CGM. 
This must be combated on both the observational and numerical fronts.
First, observations are required to better constrain the gas distribution as a function of temperature in the CGM.  In particular, in this paper we have shown that more sightlines in the central regions (within $0.5 \, \rvir$) will be powerful tools to discriminate between different feedback models.  Thus far, published works using the simulations presented here have argued that they find strong agreement with observations \citep[e.g.,][]{Fielding+17, Nelson+18, Li+20a}.  Because we have found strong differences in the extent of cold gas, the challenge for observers is clear: constrain the extent of cold gas based on stellar mass and star formation rate. 

Second, simulations must move away from tuned sub-grid models that ignore the physics in unresolved regions to physically-motivated prescriptions that are tightly bound to small-scale simulations. In upcoming SMAUG papers we will introduce our efforts to design a predictive sub-grid model for the launching and interaction of galactic winds. Our model is being built using a bottom-up approach that is based on the detailed characterization of the multiphase outflow launching properties in the high-resolution, local ISM patch simulations using the TIGRESS framework (C.-G. Kim et al. 2020 in preparation). By comparing idealized CGM simulations with a physically-motivated wind launching model and observational constraints for the inner CGM, we will be able to identify potentially important missing physical processes and develop a more complete understanding of the role of feedback and the CGM in galaxy formation.

\acknowledgements 
D.B.F. thanks Lucy Schultz for her hospitality and many useful suggestions during the preparation of this manuscript. We thank the rest of the SMAUG team for useful comments and suggestions.
This work was carried out as part of the SMAUG project. SMAUG gratefully acknowledges support from the Center for Computational Astrophysics at the Flatiron Institute, which is supported by the Simons Foundation
G.L.B. acknowledges financial support from the NSF (grant AST-1615955, OAC-1835509), and NASA (grant NNX15AB20G), and computing support from NSF XSEDE. C.-G.K. was supported in part by NASA ATP grant No. NNX17AG26G. The data used in this work were, in part, hosted on facilities supported by the Scientific Computing Core at the Flatiron Institute, a division of the Simons Foundation. E.E.S. is supported by NASA through Hubble Fellowship grant \#HF-51397.001-A awarded by the Space Telescope Science Institute, which is operated by the Association of Universities for Research in Astronomy, Inc., for NASA, under contract NAS 5-26555.

\bibliography{references}

\end{document}